\documentclass[prd,aps,twocolumn,showpacs,preprintnumbers,amsmath,amssymb]{revtex4}
\usepackage[dvips]{graphicx}
\usepackage{epsf}
\usepackage{amsmath}
\usepackage{amssymb}

\usepackage{graphicx}% Include figure files
\usepackage{dcolumn}% Align table columns on decimal point
\usepackage{bm}% bold math
\def\be{\begin{equation}}
\def\ee{\end{equation}}
\def\bea{\begin{eqnarray}}
\def\eea{\end{eqnarray}}

\DeclareMathAlphabet{\mathpzc}{OT1}{pzc}{m}{it}

\usepackage{bbm,amssymb,mathrsfs}

\begin{document}

\title{Analytical Study of Mode Coupling in Hybrid Inflation}

\author{Laurence Perreault Levasseur, Guillaume Laporte and Robert Brandenberger}

\affiliation{Department of Physics, McGill University, Montr\'eal, QC, H3A 2T8, Canada}


\begin{abstract}

We provide an analytical study of the coupling of short and long wavelength fluctuation modes 
during the initial phase of reheating in two field models
like hybrid inflation. In these models, there is - at linear order in perturbation
theory - an instability in the entropy modes of cosmological perturbations which, if
not cut off, could lead to curvature fluctuations which exceed the current observational
values. Here, we demonstrate that the back-reaction of short wavelength fluctuations
is too weak to lead to a truncation of the instability for the long wavelength modes on time 
scales comparable to the typical instability time scale of the long wavelength entropy
modes. Hence, unless there are other mechanisms which truncate the
instability, then in models such as hybrid inflation the curvature perturbations produced
during reheating on scales of current observational interest may be very important.
%%RB Major change in the abstract
 
\end{abstract}

\maketitle

%\begin{multicols}{2}{
%\small

\section{Introduction}

The inflationary scenario \cite{Guth} is the current paradigm of early universe cosmology. It addresses
several conceptual problems which Standard Big Bang cosmology was unable to solve, and - perhaps
more importantly - provides a causal mechanism to generate the primordial cosmological fluctuations
which lead to the large-scale structure and cosmic microwave background anisotropies which are
currently observed \cite{ChibMukh}. Reheating after the period of inflation is a key aspect of
inflationary cosmology (see the recent review of \cite{Reheatingrev}). During reheating the
energy is transformed from the inflaton, the scalar matter field which is responsible for providing the
inflationary expansion of space, to the matter we see today. Without reheating, the inflationary
universe would leave behind a large universe empty of any matter, in obvious contradiction with
observations. Reheating requires some coupling between the inflaton and regular matter. 

The initial studies of reheating were based on first order perturbation theory applied to the
inflaton decay \cite{initial}. This approach, however, fails to take into account the coherent
nature of the inflaton field. At the end of the period of inflation, the inflaton begins coherent
homogeneous oscillations about the ground state of its potential. In \cite{TB} (see also
\cite{DK,STB}) it was realized that this gives rise to a potential parametric instability in any
matter field which couples to the inflaton. This instability can take various forms depending
on the nature of the coupling between the inflaton and the matter field. Rather generically,
the process leads to a rapid energy transfer from the inflaton to the matter fields \footnote{Rapid
meaning on a time scale less than the Hubble time at the end of the period of inflation.}.
In a wide class of models, the resonance is of ``broad resonance" \cite{KLS1} type and affects all
fluctuations with wavenumbers smaller than a characteristic mass set by the particle
physics Lagrangian. In some cases - in particular in the hybrid inflation model which we
study in this paper - there is a tachyonic instability in the matter sector at the end of inflation,
in which case the reheating instability is of ``tachyonic resonance" \cite{tachyonic} type and
hence extremely efficient.

The inflaton field obviously couples to gravity, and therefore, as first conjectured in \cite{BKM1},
it is possible that its oscillations will lead to a
parametric resonance instability of the cosmological fluctuation modes. Due to the exponential 
growth of the causal horizon during the inflationary phase with constant Hubble radius, 
even super-Hubble modes can be causally affected by this instability \cite{FB1}. This will inherently 
cause a long wavelength curvature mode to develop which potentially exhibits a tachyonic 
resonance. These large scale modes are of prime importance, since they correspond to scales 
observable today. If an inflationary model fails to control this possible instability of the curvature 
mode, the observed smallness of the magnitude of curvature perturbations today might rule out this 
model. 

If matter consists of a single scalar field, then - as studied in
\cite{FB1} and \cite{Afshordi} - there is no parametric instability of super-Hubble scale
modes. However, if entropy fluctuations are present, then an instability of entropy modes
can occur which will lead to rapid growth of the curvature fluctuations. For this
instability to be effective, the entropy mode cannot have been exponentially redshifted
during the period of inflation. Models which satisfy the conditions for parametric instability
of the metric fluctuations were first discussed in \cite{BaVi} and \cite{FB2}.

If the duration of the parametric instability of long wavelength curvature fluctuations were long,
then the curvature fluctuations induced by the resonant entropy modes could become larger
than the primordial curvature perturbations and - in fact - larger than the observed amplitude
of the inhomogeneities. Thus, one might potentially be able to use the parametric resonant
instability of the entropy modes of the cosmological fluctuations to rule out large classes of
multi-field inflationary models.

One class of inflationary models which at linear order in perturbation theory leads to an
instability of the entropy fluctuations is hybrid inflation \cite{hybrid}. In this scenario,
two scalar fields $\phi$ and $\psi$, are involved in the inflationary process, the
slowly rolling inflaton field $\phi$ and the ``waterfall" field $\psi$.
The key point of this class of models is that while $\phi$ is slowly rolling during inflation, 
the energy density of the Universe is dominated by the potential of $\psi$. 
Over the past few years, a lot of interest has been devoted to these models, mainly since they 
provide a framework for the realization of inflation in the context of both supersymmetry 
\cite{Rachel}  and string theory (see \cite{stringinflrev} for reviews and \cite{Tye} for an
original reference). Among the promising approaches stand the D-brane / antibrane inflation 
models (e.g. the D3/D7 brane inflation model \cite{Keshav} and the KKLMMT model \cite{KKLMMT}).
Though these models may resolve some of the conceptual problems from which simple
single scalar-field  driven inflation models suffer, the large number of light moduli fields 
present in string compactifications can give rise to entropy fluctuation modes, which could in 
turn enter a parametric resonance phase at early stages of reheating \cite{BaVi,FB2}.

Hybrid inflation models are characterised by symmetry breaking 
along the $\psi$ direction: The symmetry $\psi~\rightarrow-\psi$ is spontaneously
broken for field values $\phi$ smaller than a critical value $\phi_c$. Inflation takes place
while $\phi$ is slowly rolling at field values $\phi > \phi_c$. Once the inflaton crosses
the critical value, the effective potential for $\psi$ develops a tachyonic instability which triggers 
the rapid rolling of $\psi$ towards one of the ground states of its potential. This causes inflation 
to end. 

Reheating in hybrid inflation models on a fixed Friedmann background cosmology
(i.e. no cosmological perturbations) was studied in detail in \cite{Felder1,Felder2}
(see also \cite{hybridpreh}). By means of
numerical simulations it was observed that within a very short time, non-linearities on a length
scale given by the mass of the waterfall field develop and dominate the subsequent stages
of the reheating process. It is important to study how large-scale metric perturbations
evolve during reheating in these models. In previous work, linear evolution of the entropy and 
curvature modes was studied \cite{BDD,BFL} \footnote{Other work on fluctuations beyond
linear analysis in hybrid inflation model focused on non-Gaussianities \cite{FIN,Barnaby}},
%%RB Small addition
and it was shown that it is possible that an important entropy perturbation mode develops.
In these works it
was assumed that non-linear effects, e.g. the back-reaction of small-scale fluctuations, 
would unlikely have a dominant effect on large-scale fluctuation. Due to the large range
of scales (the small wavelength modes we are interested in have wavelengths comparable
to the inverse Hubble radius at the end of inflation whereas the wavelengths of
modes we are interested in for current cosmological observations are of the order of $1 {\rm mm}$,
assuming that the scale of inflation is about $10^{15} {\rm GeV}$) numerical studies do
not have the dynamical range to study this question. Instead, an analytical understanding
of the dynamics is required. 

%%RB This paragraph is completely changed
Our work demonstrates, by means
of an analytical analysis, that the back-reaction of short wavelength modes with random
phases is too weak to truncate the instability of the long wavelength 
entropy modes. We have studied the effects of short wavelength fluctuations 
both on the background inflaton and on long wavelength modes of the inflaton and
waterfall fields. Most importantly, we have shown that the back-reaction on large scale 
fluctuations of the waterfall field 
is too weak to shut off the resonance of this entropy mode one the relevant time
scale of the instability. 
This implies that hybrid inflation models of the type analyzed in this paper may suffer
from a potential ``entropy mode problem", unless there are other ways
of truncating the resonance (e.g. the exponential decrease in the initial value
of the fluctuations modes during the period of inflation - see e.g. \cite{BFL}
for a discussion) \footnote{There is another way to test these models 
observationally: hybrid inflation models typically produce topological defects such
as cosmic strings. These strings, if stable on cosmological scales as they are in field
theory models of hybrid inflation, lead to a scaling solution of strings at all late times
(see e.g. \cite{CSrevs} for reviews on cosmic strings and structure
formation). The strings, in turn, produce line discontinuities in cosmic microwave temperature
anisotropy maps \cite{KS} which can be searched for in observational maps using
edge detection algorithms such as the Canny algorithm, as recently studied in \cite{Canny}.}.
%%RB The last part of the paragraph moved to a footnote.

%%RB New paragraph
Our study of the back-reaction of fluctuations in hybrid inflation models is not the first
analytical study. For an analytical study of the back-reaction effect of
fluctuations on the background using very different techniques than the ones we use
see \cite{Katrin}. For a study of the effects of metric fluctuations on the observable
local expansion rate of the universe due to fluctuations see \cite{Ghazal}.

The structure of this paper is as follows. In Section 2, we review the class of hybrid inflation models 
which are studied here, and introduce the perturbative approach employed to study non-linear 
interactions that arise  during preheating. In Section 3, we solve the background theory, and in 
Section 4 we review the evolution of quantum fluctuations to linear order. In Section 5, we analyze 
the generation of long wavelength second order perturbation modes sourced
%%RB Major change in the text below 
by shorter wavelength first order fluctuations (summing over the contribution from all frequencies
and assuming random phases), 
for both the 
inflaton and the waterfall field. This is the leading order back-reaction effect between
short and long wavelength modes. We also study the back-reaction of the fluctuations on
the background. We find that small-scale first-order modes with random phases have a 
contribution on the evolution of large-scale modes which does not start to dominate
until long after the instability of the first order long wavelength modes has had a chance
to develop. Hence, in spite of their large phase space, the non-linear effects of the
short wavelength perturbations do not 
become dominant for the evolution of long wavelength modes during the  early stages of 
preheating.

\section{The Model and Perturbative Approach}

We focus on the category of hybrid inflation models with Lagrangian density for matter given by:
\bea \label{lag}
\mathcal{L}_m(\phi,\psi) \, &=& \, 
\frac{1}{2}\partial_\mu \phi\partial^\mu\phi +\frac{1}{2}\partial_\nu\psi\partial^\nu\psi  \\
& & \, - \frac{1}{2}m^{2}\phi^{2}-\frac{1}{4}\lambda \left(\psi^{2}-v^{2}  \right)^{2}-\frac{1}{2}g^{2}\phi^{2}\psi^{2}
\, . \nonumber
\eea
The equations of motion form a coupled system of partial differential equations
\bea \label{odephi}
\ddot{\phi} + 3H\dot{\phi}-\frac{1}{a^{2}}\nabla^{2}\phi \, &=& \,-\left(m^{2}+g^{2}\psi^{2}\right)\phi \\
	\label{odepsi}
\ddot{\psi} + 3H\dot{\psi}-\frac{1}{a^{2}}\nabla^{2}\psi \, &=& \, 
-\left(\lambda \left(\psi^{2}-v^{2}\right)+g^{2}\phi^{2}\right)\psi \, .
\eea
The Hubble parameter $H(t)$ at time $t$ is given by:
\begin{equation}
	\label{Hubbleeqn}
		H(t)^{2} \, = \, \left (  \frac{\dot{a}}{a}\right)^{2}=\frac{8\pi G}{3}\rho \, ,
\end{equation}
where $\rho$ is the energy density, and is given by:
\begin{equation}
		\rho \, = \, \frac{1}{2}\dot{\phi}^{2}+\frac{1}{2}\dot{\psi}^{2}+\frac{1}{2}a^{-2}(\nabla \phi)^{2}+\frac{1}{2}a^{-2}(\nabla \psi)^{2}+V(\phi,\psi) \, ,
\end{equation}
where the potential $V$ is given in the second line of (\ref{lag}).
The turnover value of $\phi$ at which $\psi$ develops a tachyonic instability will then be 
\be \label{phicrit}
\phi_c \, = \, \frac{v\lambda^{1/2}}{g} \, .
\ee

To study this model we work in discrete Fourier space (discrete because of a finite
volume cutoff) and expand to second order about a homogeneous and isotropic
cosmological background. The expansion parameter $\varepsilon$ is the amplitude of the linear
fluctuations. Our goal is to study how first order perturbations of 
high $k$ modes (with wavelengths comparable to the Hubble length at the end of
inflation or to the wavelength associated with the mass of the waterfall field, whichever
is smaller) influence the low $k$ modes (modes which affect cosmological
observations today which correspond to  a scale of roughly $l_0\sim1~mm$ at the
end of inflation) at second order. We want to estimate the time interval
$\Delta t$ it will take before this back-reaction
effect becomes dominant, and we also want to see whether back-reaction effects on 
modes  with wavelength of the order $l_0$ decreases with the wavelength.

The expansion of the fields to second order in $\varepsilon$ is: 
\bea
\label{ansatz1}
\phi(x,t) \, &=& \,  \phi^{(0)}(t) + \varepsilon \delta \phi^{(1)}(x,t) + \varepsilon^2 \delta \phi^{(2)}(x,t) \\
\label{ansatz2}
\psi(x,t) \, &=& \,  0 + \varepsilon \delta \psi^{(1)}(x,t) + \varepsilon^2 \delta \psi^{(2)}(x,t) \, .
\eea
Since we will eventually be evaluating mode sums numerically, it is useful to work with
real Fourier modes. Hence, we can expand the first and second order field perturbations
as follows:
\begin{displaymath}
\delta \phi^{(1)} = \left(\sum_{n=0}^{\infty} \left[\phi^{(1)s}_{n}(t)sin\left(\frac{n\pi x}{L}\right)+\phi^{(1)c}_{n}(t)cos\left(\frac{n\pi x}{L} \right) \right] \right)
\end{displaymath}
\begin{displaymath}
 \delta \phi^{(2)} = \left(\sum_{n=0}^{\infty} \left[\phi^{(2)s}_{n}(t)sin\left(\frac{n\pi x}{L}\right)+\phi^{(2)c}_{n}(t)cos\left(\frac{n\pi x}{L} \right) \right] \right)
\end{displaymath}
\begin{displaymath}
 \delta \psi^{(1)} = \left(\sum_{n=0}^{\infty} \left[\psi^{(1)s}_{n}(t)sin\left(\frac{n\pi x}{L}\right)+\psi^{(1)c}_{n}(t)cos\left(\frac{n\pi x}{L} \right) \right] \right)
\end{displaymath}
\begin{displaymath}
 \delta \psi^{(2)} = \left(\sum_{n=0}^{\infty} \left[\psi^{(2)s}_{n}(t)sin\left(\frac{n\pi x}{L}\right)+\psi^{(2)c}_{n}(t)cos\left(\frac{n\pi x}{L} \right) \right] \right) \, .
\end{displaymath}
Here, $2L$ is the size of the finite one-dimensional box inside of which we are performing the 
discrete Fourier expansion. Physically, we need to take $L$ larger  (but not much larger)
than the largest scale of the problem we study; hence we fix it to be $\sim 1mm=6\times10^{31}l_p$. 
This effectively imposes a cutoff on the largest scale studied. For clarity, we only write explicitly 
the one-dimensional expansions, but generalisation to three-dimensional Fourier series is 
straightforward and will not modify in a crucial way the form of the obtained solutions, unless 
otherwise mentioned.

\section{Zeroth Order Expansion}

Inserting the above ansatz into the equations of motion of the system and expanding to 
zeroth order in $\varepsilon$, (\ref{odephi}) and (\ref{odepsi}) reduce to 
\bea
\ddot{\phi}^{(0)}(t) + 3H\dot{\phi}^{(0)}(t) \, &=& \, -m^{2}\phi^{(0)}(t) \\
\psi^{(0)}(t) \, &=& \, 0 \, .
\eea

For values of $|\phi|$ smaller than $\phi_c$, there is an instability of the background solution
for $\psi$. Because of this instability, the $\psi$ field grows fast on the scale of a Hubble 
expansion time. Hence, we expect the Hubble damping term to be negligible, and thus we can 
set $H\approx0$.  This amounts to setting $a(t)$ to a constant (which we pick to be 1) 
and $\dot{a}(t)=0.$ The linear equation for $\phi$ then becomes that of a harmonic oscillator 
and has the solution
\be
\phi^{(0)}(t) \, = \, A^{(0)}cos(mt) + B^{(0)}sin(mt) \, , 
\ee
where $A^{(0)}$ and $B^{(0)}$ are constants depending on the initial conditions on 
$\phi^{(0)}(t=0)$ and its derivative. Since in our case, we are interrested in the end of 
the inflationary era, we want $\phi^{(0)}(t=0)=\phi_c$ and we also want $\phi$ to be initially 
slowly rolling, i.e. $\left| \dot{\phi}^{(0)}(t=0) \right| \ll 1$. Thus, we set $B^{(0)}$ to zero and obtain:
\bea
\phi^{(0)}(t) \, &=& \, \phi_c cos(mt) \\
\psi^{(0)}(t) \, &=& \, 0 \, .
\eea

\section{First Order Expansion}

\subsection{Equations}

Now, going back to the system (\ref{odephi}) and (\ref{odepsi}) and inserting the ansatz (\ref{ansatz1}) and (\ref{ansatz2}), we expand and keep terms of first order in $\varepsilon$:
\bea
\label{1storderrealspacephi}
 \delta \ddot{\phi}^{(1)}(x,t)  &+& 3H \delta \dot{\phi}^{(1)}(x,t)-\frac{1}{a^{2}}\nabla^{2} \delta \phi^{(1)}(x,t) 
 \nonumber \\
 &=& \, -  m^{2}  \delta \phi^{(1)}(x,t) \\
 \delta \ddot{\psi}^{(1)}(x,t)  &+& 3H \delta \dot{\psi}^{(1)}(x,t)-\frac{ 1}{a^{2}}\nabla^{2} \delta \psi^{(1)}(x,t) 
 \label{1storderrealspacepsi} \\
 &=& \lambda v^{2}  \delta \psi^{(1)}(x,t)-g^2\left(\phi^{(0)}(x,t)\right)^2\left(\delta\psi^{(1)(x,t)}\right)
 \, . \nonumber
\eea

Inserting the explicit form of the first order perturbations, we make use of the orthogonality 
relations for trigonometric functions to convert (\ref{1storderrealspacephi}) and 
(\ref{1storderrealspacepsi}) to discrete Fourier space.
%%
%\begin{displaymath}
%\int^{L}_{-L}sin\left(\frac{a\pi x}{L}\right)sin\left(\frac{b\pi x}{L}\right)dx \, = \, 
%\left\{ \begin{array}  & L \delta_{ab}  \qquad if~a,~b \neq 0 \\
%0 \qquad if~a,~b=0 \end{array}\right. .
%\end{displaymath}
%%
%\begin{displaymath}
%\int^{L}_{-L}cos\left(\frac{a\pi x}{L}\right)cos\left(\frac{b\pi x}{L}\right)dx=\left\{ \begin{array} & L \delta_{ab} \qquad if~a,~b\neq 0 \\
 %\delta_{ab} 2L \qquad if~a,~b=0 \end{array}\right. .
%\end{displaymath}
%%
%\begin{equation}
%\label{orthorelations}
%\int^{L}_{-L}cos\left(\frac{a\pi x}{L}\right)sin\left(\frac{b\pi x}{L}\right)dx=0 \quad \forall ~ a,b \, .
%\end{equation}

Doing so, we obtain the following set of differential equations for the first order correction to each Fourier mode:
\bea
\ddot{\phi}^{(1)s,c}_n &+& 3H\dot{\phi}^{(1)s,c}_n + \left(\left(\frac{n\pi}{aL}\right)^2+m^2\right)\phi^{(1)s,c}_n
\nonumber \\
&=& \, 0 \\
\ddot{\psi}^{(1)s,c}_n &+& 3H\dot{\psi}^{(1)s,c}_n + \left(\left(\frac{n\pi}{aL}\right)^2-\lambda v^2+g^2\left( \phi^{(0)}\right)^2\right)\psi^{(1)s,c}_n
\nonumber \\
&=& \, 0 \, .
\eea
Replacing the wavenumber $n$ by its vectorial expression $\mathbbmtt{n}$ yields, without 
any other modification, the generalisation of the 1+1-dimensional equations to 
the 3+1-dimensional setting.

\subsection{ Solutions}

Again setting $H\approx0$, the first order $\phi$ equations become that of harmonic oscillators, and so can be solved easily:
\bea \label{1stordermodelongsol}
\phi^{(1)s,c}_{n} \, &=& \, A^{(1)s,c}_{n}cos\left( \sqrt{\left( \frac{n\pi}{aL}\right)^2+m^2}t \right) \nonumber \\
& & + B^{(1)s,c}_{n}sin\left( \sqrt{\left( \frac{n\pi}{aL}\right)^2+m^2} t\right) \, .
\eea

Let us now have a closer look at the characteristic parameter values. We have in mind a
hybrid inflation model stemming from string scale physics. Hence, the value of $v$
will be taken to be $v = 10^{-2}$ in Planck units. We choose coupling
constants  $g=10^{-2}$ and $\lambda=10^{-4}$, so that $\phi_c=10^{-2}$ (again in Planck
units).  We will consider the range $m~\epsilon~[10^{-7},10^{-5}]$ (in Planck units). 
For a value of $m$ at the upper end of this interval, the hybrid inflation model will
result in cosmological fluctuations of the observed order of magnitude (see e.g.
\cite{MFB} for a review of the theory of cosmological perturbations and \cite{RHBrev2}
for an introductory overview). We explore a range of masses $m$ in order to study
how the strength of our back-reaction effect depends on the model parameters.

Moreover, we are obviously interested in modes whose wavelength is larger
than $m^{-1}$ and larger than the Hubble radius. For such modes
$\left( \frac{n\pi}{aL}\right) \ll m$. Therefore, (\ref{1stordermodelongsol}) can be 
approximated by:
\begin{equation}
\phi^{(1)s,c}_{n} \, = \, A^{(1)s,c}_{n}cos\left( mt \right)  + B^{(1)s,c}_{n}sin\left(m t\right),
\end{equation}
which describes stable harmonic oscillation. No instability is manifest in the $\phi$ field.

On the other hand, the first order $\psi$ equations become:
\begin{equation}
\ddot{\psi}^{(1)s,c}_n(t) + \left[\left(\frac{n\pi}{aL}\right)^2 - \lambda v^2 + g^2\phi_c^2 cos^2(mt)\right]\psi^{(1)s,c}_n(t) \, = \, 0 \, .
\end{equation}
The equation 
can be reduced to the Mathieu equation by performing the transformation $z=mt$ and 
using the identity $cos^2(z)=1/2\left(1+cos(2z)\right)$:
\begin{equation}
\psi''^{(1)s,c}_n + \left[\left(\frac{n\pi}{amL}\right)^2 - \frac{\lambda v^2}{m^2} + \frac{g^2\phi_c^2}{2m^2}\left(1+ cos(2z)\right)\right]\psi^{(1)s,c}_n \, = \, 0 \, ,
\end{equation}
where the prime refers to a derivation with respect to $z$. Defining: 
\bea
q \, &=& \, \frac{-g^2\phi_c^2}{4m^2}, \\
\omega_n \, &=& \, \left(\frac{n\pi}{amL}\right)^2 - \frac{\lambda v^2}{m^2} + 2q \, , 
\eea
we indeed recover the canonical form of the Mathieu equation (see e.g. \cite{Mathieu})
\begin{equation}
\label{mathieuorder1psi}
	\psi''^{(1)s,c}_n(t) + \left[\omega_n - 2q  cos(2z)\right]\psi^{(1)s,c}_n(t) \, = \, 0 \, .
\end{equation}
The paramter $q$ is called the ``Floquet exponent", and we will call $\omega_n$ 
the ``square frequency".

For the parameter values we are using, the value of $q$ is much larger than $1$.
Hence, we are in the parameter region of either ``tachyonic resonance"
(if the tachyonic term in the expression for $\omega_n$  dominates over the third
term, or ``broad parametric resonance" if the third term dominates (the
first term in $\omega_n$ is negligible for the modes we are interested in).
In either case, all of the infrared modes which we study here will experience
an exponential instability with a growth rate characterized by the Floquet exponent. 
To first order in perturbation theory, every $\psi$ mode evolves in an independent way 
and there is no interaction between different modes. In particular, a mode that is not 
initially excited will not grow to first order at any later time (obviously, we expect
quantum vacuum fluctuations on all scales to seed the instability). Inserting the
expression (\ref{phicrit}) for $\phi_c$, we find that for all values of $n$ of interest to us
\bea
\left(\frac{n\pi}{amL}\right)^2 \, &\ll& \, \left| -\frac{\lambda v^2}{m^2} + \frac{g^2\phi_c^2}{2m^2}\right| 
\nonumber \\
&=& \, \left| - \frac{g^2\phi_c^2}{2m^2} \right| \nonumber
\eea
(independently of the value of $m$); which means that $\omega_n$ is always negative.  Hence, we conclude that  all modes we are interested in undergo tachyonic parametric resonance.

The solution to the second order differential equation (\ref{mathieuorder1psi}) can 
be written in terms of two linearly independent solutions, the so-called Mathieu 
functions $M_{ath}C$ and $M_{ath}S$:
\begin{equation}
\psi^{(1)s,c}_n(z) \, = \, C^{(1)s,c}_{n}M_{ath}C(\omega_n, q, z) + D^{(1)s,c}_{n}M_{ath}S(\omega_n, q, z) \, ,
\end{equation}
where the C's and $D's$ are the coefficients.

Note that an important property of the solution to (\ref{mathieuorder1psi}) for any choice of the
parameters is the existence of an exponential instability of parametric resonance type for 
certain ranges of the parameter value $\omega_n$. For values of $\omega_n$ falling within the 
instability range,  $\phi^{(1)s,c}_n(t)$ increa-
%\begin{widetext}

\begin{figure}[htbp]
\includegraphics[scale=0.45]{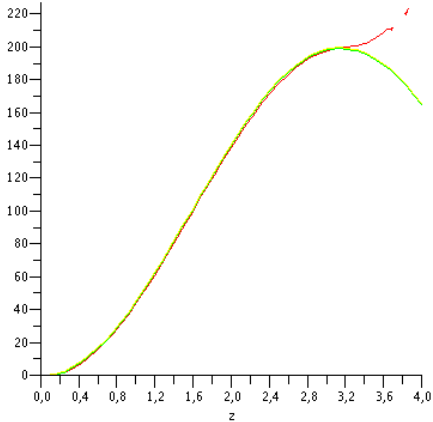}
\caption{Analytical solution to (\ref{mathieuorder1psi}) in red,  
$cosh\left(\frac{g\phi_c}{m}\left( 1-cos(ms) \right) \right)$ in green, and 
$exp\left[ \frac{g\phi_c}{m} (1-cos(ms))\right]$ in yellow, all on a logarithmic scale, 
as a function of $z=ms$; for $m=10^{-6}$ and for the reference parameters cited 
in the text. The $cosh$ function is a better approximation to the MathieuC function, 
and it is an upper bound for $z<\pi$.}
%\label{bgfig}
\end{figure}

%\end{widetext}

\noindent ses exponentially. In the
cases of tachyonic and broad parametric resonance 
\be
\phi^{(1)s,c}_n(t) \, \propto \, exp(\mu z)
\ee
for some constant $\mu$ depending on $q$. In the case of broad parametric resonance
\cite{KLS1}, the parameter $\mu$ is of the order (but slightly smaller) than 1.

%\EPSFIG[scale=0.45]{approxMathieuCsmall.eps}{\scriptsize Analytical solution to (\ref{mathieuorder1psi}) in red,  $cosh\left(\frac{g\phi_c}{m}\left( 1-cos(ms) \right) \right)$ in green, and $exp\left[ \frac{g\phi_c}{m} (1-cos(ms))\right]$ in yellow, all in logarithmic scale, as a function of $z=ms$; for $m=10^{-6}$ and the reference parameter cited in the text. The $cosh$ function is a better approximation to the MathieuC function, and it's still an upper bound for $z<\pi$.}

We can find an approximate solution of (\ref{mathieuorder1psi}) valid over the first half
of the  period of $\phi^{(0)}$ under the assumption that the initial value is $1$ (this
is a normalization) and that the initial velocity of the
mode vanishes (which, as discussed in the next subsection, is a good assumption for the
modes we are interested in). Our approximate solution in fact gives an upper bound on the
value of the mode function (which is within a factor two of the exact solution) and is given
by
\begin{equation}
\label{MathCapprox}
M_{ath}C\left(-\frac{g^2\phi_c^2}{2m^2},\frac{-g^2\phi_c^2}{4m^2},z\right) \, 
\lesssim \, cosh\left(\frac{g\phi_c}{m}(1-cos(z))\right) \, .
\end{equation} 
This result is found by approximating the equation of motion as 
\be
\psi'' \, = \, \left(\frac{g\phi_c}{m}\right)^2(1-\cos^2 z)\psi \, = \, \left(\frac{g\phi_c}{m}\right)^2\sin^2 (z) \psi \, ,
\ee
and imposing the initial conditions mentioned above. Indeed, if we set 
\be
\tilde{\psi} \, \equiv \, \cosh\left(\frac{g\phi_c}{m}(1-\cos z)\right) \, , 
\ee
we obtain 
\be
\tilde{\psi}'' \, = \, \left(\frac{g\phi_c}{m}\right)^2\sin^2 z \tilde{\psi}
+ \frac{g\phi_c}{m}\sin z\sinh\left(\frac{g \phi_c}{m^2}(1-\cos z)\right) \, . \nonumber
\ee
But $\frac{g\phi_c}{m}\sim100$ from our choice of parameters, and $\sinh x<\cosh x$. So we have 
$\tilde{\psi}>\psi$ at least from $z=0$ to $z=\pi$ (in which region $\cos z>0$).

\subsection{Initial Conditions}

Initial conditions on the first order fluctuation modes are given by the quantum fluctuations at the 
end of inflation. These modes begin on sub-Hubble scales at the beginning of the inflationary
phase in their quantum vacuum state. As reviewed e.g. in \cite{MFB,RHBrev2}, the
fluctuations freeze out and undergo squeezing once the wavelength exits the Hubble radius.
The squeezing implies that the velocity of the mode functions 
will redshift. For purely notational simplicity we have chosen to excite only 
the $\phi^{(1)c}$ and  $\psi^{(1)c}$ modes and to set the $\phi^{(1)s}$, $\psi^{(1)s}$ modes to zero. 
This implies that we are taking correlated phases for the first order modes. At the end
of the calculation we will restore the randomness of the phases and comment on the
effect that this has on the strength of the back-reaction.
%%RB Sentence added above
Because of the squeezing of the super-Hubble modes discussed above, we
start $\phi^{(1)c}$ and $\psi^{(1)c}$ with zero velocity, that is to say, we set 
$B^{(1)c}_{n}=D^{(1)c}_{n}=0~\forall~n$, while $A^{(1)c}_{n}$ and $C^{(1)c}_{n}$ are 
determined by the Bunch-Davies state. 

When numerically computing mode sums later on in this article, it is important to know the
initial amplitude of the mode functions in discrete momentum space.
In continuous Fourier space, these amplitudes are given by (in $d$ spatial dimensions):
\be
\tilde{\phi}^{(1)c}(\mathbbmtt{k}) \, = \, k^{-1/2} \, ,
\ee
where we are using the Fourier decomposition in the form
\begin{equation}
\label{continuumexptrouverdim}
	\delta\phi^{(1)}(\mathbbmtt{x},0) \, = \, \mathfrak{Re}\left[\int \frac{d^dk}{(2\pi)^d}e^{i\pi \mathbbmtt{k}  \mathbbmtt{x}}\tilde{\phi}^{(1)c}_C(\mathbbmtt{k})V^{1/2}\right] \, ,
\end{equation}
where $V$ is the spatial volume. We want to match this set of initial conditions with our 
discrete Fourier series:
\begin{equation}
\label{discreteexptrouverdim}
	\delta\phi^{(1)}(\mathbbmtt{x},0) \, = \, 
	\mathfrak{Re}\left[\sum_{\mathbbmtt{n}=0}^{\infty}e^{i\mathbbmtt{n} \mathbbmtt{x}/L}\tilde{\phi}_D^{(1)c}(\mathbbmtt{n})\right] \, .
\end{equation}

We know $\mathbbmtt{k}L=\pi\mathbbmtt{n}$ and $\Delta k L=\pi$, and thus want 
to relate $\tilde{\phi}_D^{(1)c}(\mathbbmtt{n})$ to $\tilde{\phi}^{(1)c}(\mathbbmtt{k})$ in 
terms of $\mathbbmtt{n}$. To do so, we make use of the identity (recalling that $V=(2L)^d$)
\begin{equation}
	\mathfrak{Re}\left[\frac{1}{V}\sum_{\mathbbmtt{n}=0}^\infty\right]=\mathfrak{Re}\left[\sum_{\mathbbmtt{n}=0}^\infty\left(\frac{\Delta k}{2\pi}\right)^d\right] \rightarrow \mathfrak{Re}\left[\int\frac{d^dk}{(2\pi)^d}\right]
\end{equation}
in the continuum limit in $d$ dimensions. In this limit, the discrete expansion 
(\ref{discreteexptrouverdim}) needs to converge to (\ref{continuumexptrouverdim}), i.e.:
\bea
	&&\mathfrak{Re}\left[\sum_{n=0}^{\infty} \left(\frac{\Delta k}{2\pi}\right)^d e^{\frac{i\mathbbmtt{n}\pi \mathbbmtt{x}}{L}}\tilde{\phi}_D^{(1)c}(\mathbbmtt{n})\left(\frac{2\pi}{\Delta k}\right)^d\right] \nonumber \\
	 &\rightarrow& \,  \mathfrak{Re}\left[\int\frac{d^dk}{(2\pi)^d} e^{i\mathbbmtt{k} \mathbbmtt{x}}\tilde{\phi}_D^{(1)c}(\mathbbmtt{k})\left(\frac{2\pi}{\Delta k}\right)^d\right] \\
	 &=& \, \mathfrak{Re}\left[\int \frac{d^3k}{(2\pi)^d}e^{i\mathbbmtt{k}\mathbbmtt{x}}\tilde{\phi}^{(1)c}_C(\mathbbmtt{k})V^{1/2}\right]
\, , \nonumber
\eea
where the last step expresses our requirement of convergence. Hence, for the initial
values of the discrete Fourier modes we find the relation:
\bea
	\tilde{\phi}_D^{(1)c}(\mathbbmtt{k}, 0) \, &=& \, 
	\left(\frac{\Delta k}{2\pi}\right)^d\tilde{\phi}^{(1)c}_C(\mathbbmtt{k}, 0)\left(2L\right)^{d/2}
	\nonumber \\
	&=& \, \left(\frac{1}{2L}\right)^{\frac{d-1}{2}}\frac{1}{(2\pi)^{1/2}}\frac{1}{\mathbbmtt{n}^{1/2}}  .
\eea

However, even though initial velocities of all first order modes must be zero due to freezing outside the Hubble radius, their relative phases must, in general, be random. Thus, the initial conditions must include these random phases. When measuring a first order mode, the expectation value of its amplitude in absolute value needs to be taken, which will divide the obtained amplitude by 2, while for further calculations, keeping the phase general as a random variable will be required. 

Bringing everything together, we thus write the solution for the first order $\phi$ and $\psi$ modes:
\bea
\label{sol1stphi}
	\phi^{(1)c}_{\mathbbmtt{n}}(t)  &=&  \frac{\cos(\theta_n)}{(2L)^{\frac{d-1}{2}}\sqrt{2\pi\mathbbmtt{n}}}\cos\left( mt \right) \, , \\
	 \phi^{(1)s}_{\mathbbmtt{n}}(t)  &=&  0 \, , \nonumber \\
\label{sol1stpsi}
	\psi^{(1)c}_{\mathbbmtt{n}}(t)  &=&  \frac{\cos(\theta_n)}{\left(2L\right)^{\frac{d-1}{2}}\sqrt{2\pi\mathbbmtt{n}}}\cosh\left[\frac{g\phi_c(1-\cos mt)}{m}\right] \, , \\
	 \psi^{(1)s}_{\mathbbmtt{n}}(t)  &=&  0 \, .\nonumber
\eea

\section{Second Order Expansion}

\subsection{Equations}

Going back to the system (\ref{odephi}) and (\ref{odepsi}) and again inserting the ansatz (\ref{ansatz1}) and (\ref{ansatz2}), we expand and now keep terms of second order in $\varepsilon$. We obtain
\bea \label{2storderrealspacephi}
&& \delta  \ddot{\phi}^{(2)}(x,t)  + 3H \delta \dot{\phi}^{(2)}(x,t) - \frac{ 1}{a^{2}}\nabla^{2} \delta \phi^{(2)}(x,t) \\
&& \,  = \, -  m^{2}  \delta \phi^{(2)}(x,t)-g^2\left( \delta\psi^{(1)}(x,t)\right)^2 \phi^{(0)} \nonumber
\eea
and
\bea \label{2storderrealspacepsi}
&& \delta \ddot{\psi}^{(2)}(x,t)  + 3H \delta \dot{\psi}^{(2)}(x,t) - \frac{ 1}{a^{2}}\nabla^{2} \delta \psi^{(2)}(x,t) \\
&& \, =  \, \left[\lambda v^{2}-g^2\left(\phi^{(0)}\right)^2 \right] \delta \psi^{(2)}(x,t) \nonumber \\
&& \,\,\, - \, 2g^2 \delta\phi^{(1)}(x,t) \delta\psi^{(1)}(x,t) \phi^{(0)} \nonumber
\eea

Inserting the explicit form of the first and second order perturbations, we make use of the 
orthogonality relations %(\ref{orthorelations})
for trigonometric functions to convert (\ref{2storderrealspacephi}) and (\ref{2storderrealspacepsi}) to discrete Fourier space. However, this time the process is 
slightly non-trivial due to the presence of the interaction terms at this order in perturbation 
theory which give rise to mode mixing. Indeed, the last terms in equations (\ref{2storderrealspacephi}) 
and (\ref{2storderrealspacepsi}) describe how the growth of first order perturbations will source 
second order pertubations. They involve products of modes, which requires the use of 
trigonometric identities to split these terms in a way that allows the use of the canonical orthogonality 
conditions for sines and cosines. After some algebra, this yelds the following set of differential 
equations for the second order correction to each Fourier mode. In one spatial dimension 
(recalling that $\phi^{(1)s}_{0}(t)=\psi^{(1)s}_{0}(t)=0$ by definition, and  
$\phi^{(1)c}_{0}(t)=\psi^{(1)c}_{0}(t)=0$ because these modes are part of the background)
we obtain the following results:

%}
%\end{multicols}
%\small

\begin{widetext}

\noindent For $n\geq1$, the equations describing the back-reaction of the fluctuation modes
on the perturbations themselves take the form
%%:
	\begin{equation}
	\label{1stphiode}
		\centerdot~~ \ddot{\phi}^{(2)s}_{n}(t)+ 3H\dot{\phi}^{(2)s}_{n}(t) + \left(\frac{n \pi}{aL}\right)^{2} \phi^{(2)s}_{n}(t)  = 
- m^{2}\phi^{(2)s}_{n}(t)- g^{2}\left(\phi^{(0)}(t)\right) \sum_{j=1}^{\infty}\left[ \psi^{(1)s}_{j}(t) \left(\psi^{(1)c}_{|j-n|}(t) -\psi^{(1)c}_{j+n}(t)\right)   \right] 
	\end{equation}
	
		\begin{displaymath}
		\centerdot~~ \ddot{\phi}^{(2)c}_{n}(t) + 3H\dot{\phi}^{(2)c}_{n}(t) + \left(\frac{n \pi}{aL}\right)^{2}\phi^{(2)c}_{n}(t) = - m^{2} \phi^{(2)c}_{n}(t)\qquad\qquad \qquad\qquad\qquad\qquad\qquad\qquad
	\end{displaymath}
	\begin{equation}
	\label{2ndphiode}
		\qquad\qquad\qquad- \frac{g^{2}}{2} \left(\phi^{(0)}(t)\right)\sum_{j=1}^{\infty}\left[\psi^{(1)s}_{j}(t)\left(  \psi^{(1)s}_{j-n}(t)+ \psi^{(1)s}_{j+n}(t)\right) + \psi^{(1)c}_{j}(t)\left( \psi^{(1)c}_{|j-n|}(t)+ \psi^{(1)c}_{j+n}(t)\right)
		\right]
	\end{equation}
	
		\begin{displaymath}
		\centerdot~~\ddot{\psi}^{(2)s}_{n}(t) + 3H\dot{\psi}^{(2)s}_{n}(t) + \left(\frac{n \pi}{aL}\right)^{2} \psi^{(2)s}_{n}(t) = \lambda v^{2}\psi^{(2)s}_{n}(t)- g^{2}  \left(\phi^{(0)}(t)\right)^{2}\psi^{(2)s}_{n}(t)\qquad\qquad\qquad
	\end{displaymath}
	\begin{equation}
	\label{1stpsiode}
		\qquad\qquad\qquad- g^{2}\left(\phi^{(0)}(t)\right) \sum_{k=1}^{\infty} \left[  \left(\phi^{(1)c}_{|k-n|}(t)-\phi^{(1)c}_{k+n}(t)\right)\psi^{(1)s}_{k}(t)+ \left(\psi^{(1)c}_{|k-n|}(t)-\psi^{(1)c}_{k+n}(t)\right)\phi^{(1)s}_{k}(t) \right] 
	\end{equation}

\begin{displaymath}
		\centerdot ~~\ddot{\psi}^{(2)c}_{n}(t) + 3H \dot{\psi}^{(2)c}_{n}(t) +\left(\frac{n \pi}{aL}\right)^{2} \psi^{(2)c}_{n}(t)= \lambda v^{2} \psi^{(2)c}_{n}(t)  -g^{2}\left(\phi^{(0)}(t)\right)^{2}\psi^{(2)c}_{n}(t) \qquad\qquad\qquad
	\end{displaymath}
	\begin{equation}
	\label{2ndpsiode}
		\qquad\qquad\qquad- g^{2} \left(\phi^{(0)}(t)\right)\sum_{k=1}^{\infty} \left[ \psi^{(1)s}_{k}(t)\left(  \phi^{(1)s}_{k-n}(t)+ \phi^{(1)s}_{k+n}(t)\right)+\psi^{(1)c}_{k}(t)\left(   \phi^{(1)c}_{|k-n|}(t)+ \phi^{(1)c}_{k+n}(t)\right)  \right] \, ,   
	\end{equation}

\noindent while for $n=0$, that is, for the back-reaction  on the 
background fields, we have:
	\begin{equation} 
	\label{2ndphioden0}
		\centerdot~~\ddot{\phi}^{(2)c}_{0}(t) + 3H\dot{\phi}^{(2)c}_{0}(t) = - m^{2} \phi^{(2)c}_{0}(t)- g^{2}\left(\phi^{(0)}(t)\right) \left(\sum_{j=1}^{\infty} \left[ \frac{1}{2}\left(\psi^{(1)s}_{j}(t)\right)^{2}+\frac{1}{2} \left(\psi^{(1)c}_{j}(t)\right)^{2}  \right] \right)
	\end{equation}

	\begin{equation}
	\label{2ndpsioden0}
		\centerdot~~\ddot{\psi}^{(2)c}_{0}(t)  + 3H\dot{\psi}^{(2)c}_{0}(t)  = \lambda v^{2}\psi^{(2)c}_{0}(t)- g^{2}\left(\phi^{(0)}(x,t)\right)^{2}\psi^{(2)c}_{0}(t)- g^{2} \left(\phi^{(0)}(t)\right)\sum_{j=1}^{\infty} \left[ \phi^{(1)s}_{j}(t)\psi^{(1)s}_{j}(t)+\phi^{(1)c}_{j}(t)\psi^{(1)c}_{j}(t)\right] \, .
	\end{equation}	

\end{widetext}

Note that the phases cancel out in the back-reaction on the inflaton field (in Eq. \ref{2ndphioden0}), 
but not in any of the other equations.
%%RB Above sentence added

%\begin{multicols}{2}{
%\small

The physics which these equations describe is the following:
Since the system initially has no 
second order perturbations, it is the interaction of two first order modes 
whose wavenumbers add up to $k$ that will source fluctuations of wavenumber $k$ 
at second order. The second order perturbation at wavenumber $k$ is affected
by all first order modes. Hence, even though the effect of each individual first order
mode is of the order $\varepsilon^2$, the large phase space of modes which
contribute can lead to a large back-reaction effect \footnote{Similarly in spirit,
the large phase space of linear perturbation modes can lead to a large
back-reaction effect of linear cosmological fluctuations on the background
metric, an effect studied in \cite{Abramo} and reviewed in \cite{RHBrev1}.}.

The above equations can luckily be reduced slightly. 
Since we have chosen not to excite the $\phi^{(1)s}$ and $\psi^{(1)s}$ modes, no second 
order sinusoidal fluctuations will arise, that is, $\phi^{(2)s} _n=\psi^{(2)s}_n=0$ for all $n$, and 
so there is no need to consider equations (\ref{1stphiode}) and (\ref{1stpsiode}). Moreover, 
$H$ can again be set to zero; and every term involving $\phi^{(2)s} _j$ or $\psi^{(2)s}_j$ in 
the interaction sum acting as a source in each equation can be set to zero. Also, as 
discussed above, the terms linear in the fields having as coefficient $\left(\frac{n\pi}{aL}\right)^2$ 
in equations (\ref{1stphiode}) through (\ref{2ndpsiode}) are negligible compared to the mass 
term of the fields, and thus can be dropped. However, recall that this conclusion was reached 
by imposing a cutoff equal to the Hubble radius on the smallest scale excited to first order. 
Consequently, all sums involving interactions of first order modes acting as source terms 
for the second order perturbation modes can be performed up to 
$n=\frac{LH}{\pi}=\frac{L\lambda^{1/2} v^2}{2\pi\sqrt{3}}\sim6\times10^{24}$.

Generalizing these equations from the 1+1-dimensional case to the higher-dimensional case 
this time is a bit more involved than simply replacing the wavenumber $n$ by its vectorial expression 
$\mathbbmtt{n}$. In fact, the simple replacement works for 
every terms except for the interaction sum in each equation, 
which needs to be modified as follows:

%}
%\end{multicols}
%\small

\begin{widetext}

\begin{equation}
\sum_{j_1,...,j_d=0}^{\frac{LH}{\pi}} \psi^{(1)c}_{j_1...j_d}(t)\psi^{(1)c}_{k_1...k_d}(t)\left[\frac{1}{2}\begin{cases}  \delta_{k_1, |j_1-n_1|}+ \delta_{k_1, j_1+n_1} ~~ ,~j_1\neq n_1 \\
		 2 \delta_{k_1, |j_1-n_1|}+ \delta_{k_1, j_1+n_1}  ~~ , ~j_1=n_1\end{cases} \right] ...\left[\frac{1}{2} \begin{cases}  \delta_{k_d, |j_d-n_d|}+ \delta_{k_d, j_d+n_d} ~~ ,~j_d\neq n_d \\ 
		2 \delta_{k_d, |j_d-n_d|}+ \delta_{k_d, j_d+n_d}  ~~ , ~j_1=n_d\end{cases} \right]
\end{equation}
\begin{equation}
\sum_{j_1,...,j_d=0}^{\frac{LH}{\pi}} \psi^{(1)c}_{j_1...j_d}(t)\phi^{(1)c}_{k_1...k_d}(t)\left[ \begin{cases}  \delta_{k_1, |j_1-n_1|}+ \delta_{k_1, j_1+n_1} ~~ ,~j_1\neq n_1 \\
		2 \delta_{k_1, |j_1-n_1|}+ \delta_{k_1, j_1+n_1}  ~~ , ~j_1=n_1\end{cases} \right] ...\left[ \begin{cases}  \delta_{k_d, |j_d-n_d|}+ \delta_{k_d, j_d+n_d} ~~ ,~j_d\neq n_d \\
		2 \delta_{k_d, |j_d-n_d|}+ \delta_{k_d, j_d+n_d}  ~~ , ~j_d=n_d\end{cases} \right]
\end{equation}
\begin{equation}
\sum_{j_1,...,j_d=0}^{\frac{LH}{\pi}} \psi^{(1)c}_{j_1...j_d}(t)\psi^{(1)c}_{k_1...k_d}(t)\left[\frac{1}{2} \begin{cases}  \delta_{k_1, j_1} ~~ ,~j_1\neq 0 \\
		2 \delta_{k_1, 0}  ~~ , ~j_1=0\end{cases} \right] ...\left[\frac{1}{2} \begin{cases}  \delta_{k_d, j_d} ~~ ,~j_d\neq 0 \\
		2 \delta_{k_d,0}  ~~ , ~j_d=0\end{cases} \right]
\end{equation}
\begin{equation}
\sum_{j_1,...,j_d=0}^{\frac{LH}{\pi}} \psi^{(1)c}_{j_1...j_d}(t)\phi^{(1)c}_{k_1...k_d}(t)\left[\begin{cases} \delta_{k_1, j_1} ~~ ,~j_1\neq 0 \\
		2 \delta_{k_1, 0}  ~~ , ~j_1=0\end{cases} \right] ...\left[\begin{cases} \delta_{k_d, j_d} ~~ ,~j_d\neq 0 \\
		2 \delta_{k_d,0}  ~~ , ~j_d=0\end{cases} \right]
\end{equation}
for the $\phi^{(2)c}_{\mathbbmtt{n}}$, $\psi^{(2)c}_{\mathbbmtt{n}}$, $\phi^{(2)c}_0$ and 
$\psi^{(2)c}_0$ equations, respectively, in the case of $d$ spatial dimensions. In particular, 
for $d=3$, the sum for the $\phi^{(2)c}_{n}$ equation (\ref{2ndphiode}) can be rewritten as:
\begin{displaymath}
	-\frac{g^2}{8}\phi^{(0)}\left[\sum_{i,j,k=0}^{\frac{LH}{\pi}}\psi^{(1)c}_{ijk}\psi^{(1)c}_{|i-n_x| |j-n_y| |k-n_z|}+3\sum_{j,k=0}^{\frac{LH}{\pi}}\psi^{(1)c}_{n_xjk}\psi^{(1)c}_{0|j-n_y||k-n_z|}+3\sum_{k=0}^{\frac{LH}{\pi}}\psi^{(1)c}_{n_xn_yk}\psi^{(1)c}_{0 0 |k-n_z|}+\sum_{i,j,k=0}^{\frac{LH}{\pi}}\psi^{(1)c}_{ijk}\psi^{(1)c}_{(i+n_x) (j+n_y) (k+n_z)}\right.
\end{displaymath}
\begin{displaymath}
	+3\left(\sum_{i,j,k=0}^{\frac{LH}{\pi}}\psi^{(1)c}_{ijk}\psi^{(1)c}_{|i-n_x| |j-n_y| (k+n_z)}+2\sum_{j,k=0}^{\frac{LH}{\pi}}\psi^{(1)c}_{n_xjk}\psi^{(1)c}_{0 |j-n_y| (k+n_z)}+\sum_{k=0}^{\frac{LH}{\pi}}\psi^{(1)c}_{n_xn_yk}\psi^{(1)c}_{0 0 (k+n_z)}\right)
\end{displaymath}
\begin{equation}
\label{2ndphiode3d}
	\left.+3\left(\sum_{i,j,k=0}^{\frac{LH}{\pi}}\psi^{(1)c}_{ijk}\psi^{(1)c}_{|i-n_x| (j+n_y) (k+n_z)}+\sum_{j,k=0}^{\frac{LH}{\pi}}\psi^{(1)c}_{n_xjk}\psi^{(1)c}_{0 (j+n_y) (k+n_z)}\right)\right] \, .
\end{equation}
Similarly, for the $\psi^{(2)c}_{n}$ equation (\ref{2ndpsiode}):
\begin{displaymath}
	-g^2\phi^{(0)}\left[\sum_{i,j,k=0}^{\frac{LH}{\pi}}\psi^{(1)c}_{ijk}\phi^{(1)c}_{|i-n_x| |j-n_y| |k-n_z|}+3\sum_{j,k=0}^{\frac{LH}{\pi}}\psi^{(1)c}_{n_xjk}\phi^{(1)c}_{0|j-n_y||k-n_z|}+3\sum_{k=0}^{\frac{LH}{\pi}}\psi^{(1)c}_{n_xn_yk}\phi^{(1)c}_{0 0 |k-n_z|}+\sum_{i,j,k=0}^{\frac{LH}{\pi}}\psi^{(1)c}_{ijk}\phi^{(1)c}_{(i+n_x) (j+n_y) (k+n_z)}\right.
\end{displaymath}
\begin{displaymath}
	+3\left(\sum_{i,j,k=0}^{\frac{LH}{\pi}}\psi^{(1)c}_{ijk}\phi^{(1)c}_{|i-n_x| |j-n_y| (k+n_z)}+2\sum_{j,k=0}^{\frac{LH}{\pi}}\psi^{(1)c}_{n_xjk}\phi^{(1)c}_{0 |j-n_y| (k+n_z)}+\sum_{k=0}^{\frac{LH}{\pi}}\psi^{(1)c}_{n_xn_yk}\phi^{(1)c}_{0 0 (k+n_y)}\right)
\end{displaymath}
\begin{equation}
\label{2ndpsiode3d}
	\left.+3\left(\sum_{i,j,k=0}^{\frac{LH}{\pi}}\psi^{(1)c}_{ijk}\phi^{(1)c}_{|i-n_x| (j+n_y) (k+n_z)}+\sum_{j,k=0}^{\frac{LH}{\pi}}\psi^{(1)c}_{n_xjk}\phi^{(1)c}_{0 (j+n_y) (k+n_z)}\right)\right] \, .
\end{equation}
Finally, for the $\phi^{(2)c}_{0}$ $\psi^{(2)c}_{0}$ equations (\ref{2ndphioden0}) and 
(\ref{2ndpsioden0}), the sum at the end of each equation gets replaced by, respectively:
\bea
\label{3dgenphipsi0}
	&& g^2\phi^{(0)}\left[\frac{1}{8}\sum_{i,j,k=1}^{\frac{LH}{\pi}}\left(\psi^{(1)c}_{ijk}\right)^2 +\frac{3}{4}\sum_{j,k=1}^{\frac{LH}{\pi}}\left(\psi^{(1)c}_{0jk}\right)^2+\frac{3}{2}\sum_{k=1}^{\frac{LH}{\pi}}\left(\psi^{(1)c}_{00k}\right)^2\right]  \nonumber \\
	&& g^2\phi^{(0)}\left[\sum_{i,j,k=1}^{\frac{LH}{\pi}}\psi^{(1)c}_{ijk}\phi^{(1)c}_{ijk} +6\sum_{j,k=1}^{\frac{LH}{\pi}}\psi^{(1)c}_{0jk}\phi^{(1)c}_{0jk}+12\sum_{k=1}^{\frac{LH}{\pi}}\psi^{(1)c}_{00k}\phi^{(1)c}_{00k}\right] \, .
\eea

\end{widetext}

%\begin{multicols}{2}{
%\small

%\subsection{Solutions}

From the above discussion, we see that the differential equations for the second order 
perturbation modes reduce to that of driven harmonic oscillators, with a constant period 
in the case of $\phi^{(2)c}$, and with a time-dependent period in the case of 
$\psi^{(2)c}$. It is thus possible to solve them by making use of the Green's function method.

For the $\phi^{(2)c}$ equations, we use the causal Green's function for a simple harmonic 
oscillator:
\begin{equation}
\label{phiGreenfct}
	G(s,t) \, = \,  \begin{cases}  \frac{1}{m}sin(m(t-s)) ~ ,~for~t\ge0 \\
		 0 \qquad\qquad\qquad~~ ,~for~t<0\end{cases} \, ,
\end{equation}
while for the $\psi^{(2)c}$ equation, we make use of the function:
\begin{equation}
\label{psiGreenfct}
	G(s,t) \, = \,  \begin{cases} \frac{1}{m}M_{ath}S(\omega_0,q,m(t-s)) ~ ,~for~t\ge0 \\
		 0 \qquad\qquad\qquad\qquad\qquad~~ ,~for~t<0\end{cases}
\end{equation}
where $\omega_0$ and $q$ are defined as above.

In the following, we first analyze the back-reaction effect on the background fields,
and then move on to study the back-reaction of first order fluctuations on the
perturbation modes themselves \footnote{In the case of cosmological perturbations,
the latter problem was studied in \cite{Martineau}.}

\subsection{Back-reaction on the $\phi$ Background}

We first tackle the description of the mode $\phi^{(2)c}_{\mathbbmtt{n}=0}$, whose 
evolution is dictated by equation (\ref{2ndphioden0}) generalised to 3 spatial 
dimensions. This mode describes how second order perturbations of the $\phi$ field 
will modify  how the background inflaton $\phi^{(0)}$ is oscillating around the 
minimum of its potential. From (\ref{2ndphioden0}) and (\ref{3dgenphipsi0}), we see that 
it is  the growth of the amplitude of the first order perturbations in the $\psi$ field that will 
source this back-reaction.

More precisely, solving the $\phi^{(2)c}_0$ equation using the Green's function 
method and making use of (\ref{phiGreenfct}), we obtain the following solution:
\bea
	&& \phi^{(2)c}_{0}(t) \, = \, 
	\int_0^{t}ds\frac{-1}{m}\sin(m(t-s))g^{2}\phi_c \cos(ms) \nonumber\\
	 && \times\left[\frac{1}{8}\sum_{i,j,k=1}^{\frac{LH}{\pi}}\left(\psi^{(1)c}_{ijk}\right)^2 +\frac{3}{4}\sum_{j,k=1}^{\frac{LH}{\pi}}\left(\psi^{(1)c}_{0jk}\right)^2\right. \nonumber\\
	 && \left.+\frac{3}{2}\sum_{k=1}^{\frac{LH}{\pi}}\left(\psi^{(1)c}_{00k}\right)^2 \right]  \, . 
\eea
%%
%%RB Sentence added below
Note that the phases won't cause any major cancellation of the terms in the summation, since all terms are positive. The expectation value of the sum can thus easilly be taken, which will simply add an additional factor of $1/2$ to the solution (that is, the expectation value of $\cos^2\theta$ over one period). Thus, we can write:
\bea \label{rawphi0sol}
	&&\phi^{(2)c}_{0}(t) \, = \, \frac{-g^{2}\phi_c}{4\pi m}\frac{1}{\left(2L\right)^{2}}\int_0^{t}ds\sin(m(t-s)) \cos(ms)
	\nonumber \\
	&& \times\cosh^2\left[\frac{g\phi_c(1-\cos ms)}{m}\right]\left[\frac{1}{8}\sum_{i,j,k=1}^{\frac{LH}{\pi}}\frac{1}{(i^2+j^2+k^2)^{1/2}} \right. \nonumber \\
          && \left.+\frac{3}{4}\sum_{j,k=1}^{\frac{LH}{\pi}}\frac{1}{(j^2+k^2)^{1/2}}+\frac{3}{2}\sum_{k=1}^{\frac{LH}{\pi}}\frac{1}{k} \right]  \, .
\eea
We see that the time-dependent part of the sums factors out, which means that the summations 
can be performed independently of the time integral. This process of summing is made much 
easier by replacing the sums with integrals, which is a resonnable approximation, since the 
range of the $i$, $j$, $k$ variables is quite wide. Moreover, instead of integrating over a 
cube of length size $\frac{LH}{\pi}$ in momentum space, we integrate over one eighth of a 
sphere of radius $\frac{LH}{\pi}$, which yields the result:
\bea
\label{monopoleterm}
	\sum_{i,j,k=1}^{\frac{LH}{\pi}}\frac{1}{(i^2+j^2+k^2)^{1/2}}  \, &\approx& \,
	\frac{1}{8}\int_1^{\frac{LH}{\pi}}r^2drd\Omega \frac{1}{r} \nonumber \\
	&=& \,  \frac{ (L\lambda^{1/2}v^2)^2}{48\pi} \, .
\eea
Similarly,
\begin{equation}
	\sum_{j,k=1}^{\frac{LH}{\pi}}\frac{1}{(j^2+k^2)^{1/2}}  \, \approx \, 
	\frac{1}{4}\int_1^{\frac{LH}{\pi}}rdrd\theta \frac{1}{r}=  \frac{L\lambda^{1/2}v^2}{4\sqrt{3}}
\end{equation}
and
\begin{equation}
	\sum_{k=1}^{\frac{LH}{\pi}}\frac{1}{k} \,  \approx \,
	\int_1^{\frac{LH}{\pi}}dk \frac{1}{k}=ln\left(\frac{L\lambda^{1/2}v^2}{2\sqrt{3}\pi}\right) \, .
\end{equation}
These sums are dominated by the small wavelength first order modes 
since these modes have the largest volume in momentum space. Hence, 
since the momentum space is three-dimensional, it is clear that the first 
of the sums dominates over the two other. Indeed, for $m=10^{-6}$ (in
Planck mass units), the first term contributes $\mathcal{O}(10^{49})$, 
compared to $\mathcal{O}(10^{25})$ and $\mathcal{O}(10)$ for the two others. 
The prefactor of the integral (\ref{rawphi0sol}) therefore becomes 
$-\frac{g^{4}\phi_c^3v^2}{2^{2}8^{3}3\pi^2m}$, independently of $L$, and the integral itself 
can be performed numerically.

As a stability and consistency check of our method, we study how the amplitude of the second 
order fluctuations in $\phi$ affects the background field as we vary the value of the mass of the 
inflaton field in Figure \ref{phi20varm}. The first striking feature of these graphs is that, for all 
considered values of $m$, the second order perturbation of the background mode becomes 
dominant before $\phi$ reaches the minimum of the potential, that is, within a short time 
relative to the time scale of the problem.

Moreover, as $m$ increases from $10^{-7}$ to $10^{-5.5}$, the fraction of the phase of $\phi^{(0)}$ 
needed before the back-reaction term starts to dominate decreases. Equivalently, the 
value of $mt$  when the amplitude of the second order perturbations becomes of similar 
order as the background field $\phi^{(0)}$ shrinks as the mass of the inflaton is 
increased. 
%This can be understood by noting that the larger the mass of the inflaton is, the 
%smaller the timescale $m^{-1}$ of the problem becomes, so the slower the evolution 
%of the inflaton will be.

However, in none of the cases studied is the time elapsed before the back-reaction 
effects on the background field becomes dominant shorter than
\be 
| m_\psi |^{-1} \, = \, \lambda^{-1/2} v^{-1} 
\ee
(which is $10^{3}$ Planck times for the parameter values we have chosen), the other relevant 
time scale of the problem under study which is the typical time scale of the
motion in the tachyonic direction. Since the tachyonic field sources the instability at the end of 
inflation, its inverse mass will set the time scale dominating the growth 

%}
%\end{multicols}
%\small
\begin{widetext}

%\EPSFIG[scale=0.27]{phipsi0compare.eps}{\scriptsize Comparison of $\phi$ background $\phi^{(0)}$ (green) with the second order back-reaction $\phi^{(2)c}_0$ (blue, top line) and with the second order back-reaction $\psi^{(2)c}_0$ (blue, bottom line), in planck units, as a function of Planck time, for values of the mass of the inflaton of $m=10^{-5.6}$, $m=10^{-6}$ and $m=10^{-6.4}$ in Planck units. $\phi$ and $\psi$ have units of mass, and are displayed in Planck units. Every plot ranges up to indentical phase of the background field $\phi^{(0)}$, that is, up to $mt=\pi/2$ Planck time, which allows for comparison of the relative growth of $\phi^{(2)c}_0$ and $\psi^{(2)c}_0$ compared to $\phi^{(0)}$. As $m$ grows, the value of $mt$ at which $\phi^{(2)c}_0$ becomes dominant over $\phi^{(0)}$ increases. Similar conclusions are drawn studying the growth of $\psi^{(2)c}_0$ as a function of $m$. For a fixed value of $m$, the growth of $\phi^{(2)c}_0$ is observed to be faster than the growth of $\psi^{(2)c}_0$, but the overall timescale over which they become dominant is the same.}{phi20varm}

\begin{figure}[htbp] 
\includegraphics[scale=0.27]{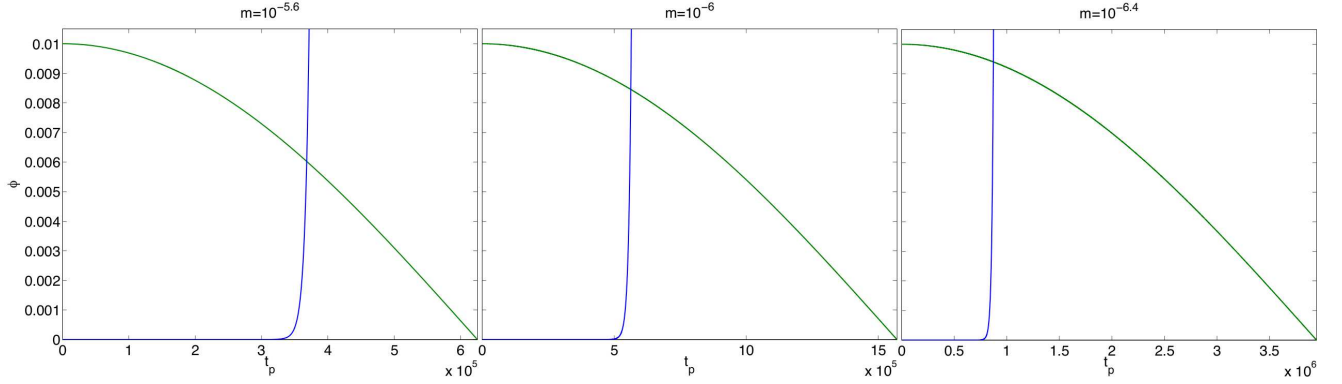}
\caption{Comparison of the $\phi$ background $\phi^{(0)}$ (green) with the second order 
back-reaction $\phi^{(2)c}_0$ (blue) as a function of Planck time, for values of 
the mass of the inflaton of $m=10^{-5.6}$, $m=10^{-6}$ and $m=10^{-6.4}$ in Planck units. 
$\phi$ has units of mass, and is displayed in Planck units. Every plot ranges up to indentical phase of the background field $\phi^{(0)}$, that is, 
up to $mt=\pi/2$, which allows for a comparison of the relative growth of $\phi^{(2)c}_0$ 
compared to $\phi^{(0)}$. As $m$ grows, the value of $mt$ at which 
$\phi^{(2)c}_0$ becomes dominant over $\phi^{(0)}$ increases.}
\label{phi20varm}
\end{figure}

\end{widetext}

%\begin{multicols}{2}{
%\small

\noindent of second order 
perturbations. Varying $m$ should thus affect the growth of second order perturbations 
in a very smooth and mild way, which is what Figure \ref{phi20varm} shows. 

\begin{figure}[] 
\includegraphics[scale=0.29]{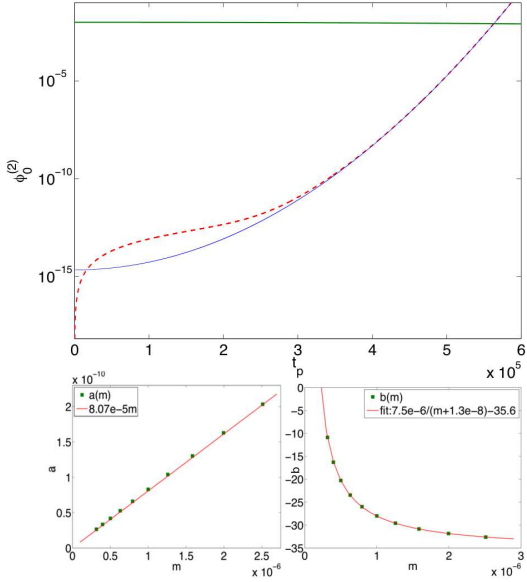}
\caption{\textsc{Top panel:} Fit of the amplitude of the second order back-reaction $|\phi^{(2)c}_0|$ (dashed red line) with the function $\sim e^{at^2+b}$ (blue line), for $m=10^{-6}$. The fitting function is 
found to be in good agreement with the growth rate of $|\phi^{(2)c}_0|$ after $mt=0.3$. The $\phi$ 
background $\phi^{(0)}$ (green) is also presented in order to provide the time scale of the 
growth of $|\phi^{(2)c}_0|$ and to provide a scale of its relative amplitude.   
\textsc{Bottom left}: Fitted value of $a(m)$ as a function of $m$ for values of $m$
ranging from $m=10^{-6.5}$ to $m=10^{-5.6}$ (green points). The fit (in red) shows that the points 
obey a linear relationship over the studied range of $m$.  \textsc{Bottom right}: Fitted 
value of $b(m)$ as a function of $m$, over the same range. The fit (in red) exhibits that 
$b(m)$ grows as $\frac{1}{m}$ over the studied values of $m$.}
 \label{fitexamplephi0}
 \end{figure}

The $\phi^{(2)c}_0$ mode is found to grow as $\sim -e^{mt^2}$, an instability that rapidly 
outruns the amplitude corresponding to the first order perturbation modes, which we found above 
to oscillate as $\sim\cos mt$. A fit of the function $-e^{at^2+b}$ with $b$ and $c$ as fitting 
parameters, presented in Figure \ref{fitexamplephi0} for the case $m=10^{-6}$, is found 
to be in agreement with the analytical values over a time scale of $m^{-1}$. However, it is to be expected that during its very early evolution, the relative error between the fit and the numerical $\phi^{(2)c}_0$ 
will be large. Indeed, $\phi^{(2)c}_0(0)=0$, while the exponential fitting function can 
never reach zero.

A plot of how the fitting parameters $a$ and $b$ vary with $m$ is displayed in the bottom 
part of Figure \ref{fitexamplephi0}, confirming the stability and smoothness of the evolution 
of the  $\phi^{(2)c}_0$ mode under variation of the inflaton's mass. The parameter $a$ goes as 
$\sim m$, while the behavior of $b$ was found to match a $\sim 1/m$ function. Therefore, we find 
the following fitting function for the growth of the second order back-reaction on the background:
\begin{equation}
	\phi^{(2)c}_0 \, \approx \,  -exp\left[8.071e\mathrm{\textendash}5 mt^2+\frac{7.477e\mathrm{\textendash}6}{(m+1.33e\mathrm{\textendash}8)}-35.60\right]
\end{equation}

\subsection{Back-reaction on the $\psi$ Background}

We now describe the mode $\psi^{(2)c}_{\mathbbmtt{n}=0}$, whose evolution is dictated by 
equation (\ref{2ndpsioden0}) generalised to 3 spatial dimensions. This mode describes how 
second order perturbations of the $\psi$ field will induce the growth of a spatially homogeneous 
contribution to the tachyonic field. At first order in perturbation theory the mode is obviously 
unexcited, and so there is no non-trivial background to compare its growth against. Note that
a second order $\psi$ background only develops since our choice of phases of the first order
fluctuations breaks the $\psi \rightarrow - \psi$ symmetry of the Lagrangian. Averaged over the
entire universe, we would expect the symmetry to be restored. However, in a finite volume there
is no reason why the phases should obey the symmetry.

Combining the $\psi^{(2)c}_0$ equation and the Green's function (\ref{psiGreenfct}), 
we obtain the following solution:
\bea
	\psi^{(2)c}_{0}(t) \, &=& \, \int_0^{t}ds\frac{(-1)}{m}M_{ath}S(\omega_0,q,m(t-s)) g^{2}\phi_c \cos(ms) 
	\nonumber \\
	&& \times \left[\sum_{i,j,k=1}^{\frac{LH}{\pi}}\psi^{(1)c}_{ijk}\phi^{(1)c}_{ijk} +6\sum_{j,k=1}^{\frac{LH}{\pi}}\psi^{(1)c}_{0jk}\phi^{(1)c}_{0jk} \right. \nonumber \\
	&& + \left. 12\sum_{k=1}^{\frac{LH}{\pi}}\psi^{(1)c}_{00k}\phi^{(1)c}_{00k}\right]
\eea
Substituting the solution for $\phi^{(1)c}_{\mathbbmtt{n}}$ and $\psi^{(1)c}_{\mathbbmtt{n}}$ 
obtained in (\ref{sol1stphi}) and (\ref{sol1stpsi}), respectively, the sums to be performed 
turn out to be the same as above for $\phi^{(2)c}_{0}$. 
%%RB Begin extra
However, there is one
important difference: in the case of the back-reaction equation for the background $\phi$
field, the phases of the first order modes cancelled out. This will not be the case here.
We will take this difference into account by performing the sums including random
phases (see Appendix). 
%%RB End extra
Moreover, the $M_{ath}S$ function 
appearing in the Green's function can be dealt with by means of the same kind of 
approximation performed for the $M_{ath}C$ function in (\ref{MathCapprox}):
\begin{equation}
	M_{ath}S\left(-\frac{g^2\phi_c^2}{2m^2},\frac{-g^2\phi_c^2}{4m^2},z\right)\lesssim \sinh\left(\frac{g\phi_c}{m}(1-\cos(z))\right)
\end{equation}
%%

%%RB Small change below
Proceeding with these substitutions and performing the sums as a random walk, we get:
\bea
	\psi^{(2)c}_{0}(t)  &=& \frac{-g^2\phi_c\lambda^{\frac{1}{4}}v}{3^{\frac{1}{4}}8\pi^2mL^{\frac{3}{2}}}\int_0^{t}ds \sinh\left[\frac{g\phi_c(1-\cos(m(t-s)))}{m}\right]  \nonumber \\
	  && \times\cos^2(ms)\cosh\left[\frac{g\phi_c(1-\cos ms)}{m}\right]
\eea
which can now be integrated numerically. The result turns out to be quite similar to the one obtained for $\phi^{(2)c}_{0}$: the second order background mode again grows as $\sim -e^{t^2}$, which is much 
faster than the first order perturbation modes, which we found above to grow as 
$\sim \cosh(1-\cos mt)$. However, due to the cancellation of terms in the sum induced by the random 
walk, their overall amplitude is significantly smaller. In fact, even their rapid growth cannot compensate 
for this factor over the time scale set by the mass of $\phi$.
%%RB part of the sentence dropped below
%and becomes comparable to the $\phi$ background on 
%timescales similar to what was found above for the $\phi^{(2)c}_{0}$ back-reaction to 
%become dominant. This similarity 

%\EPSFIG[scale=0.3]{premierfit.eps}{\scriptsize \textsc{Top part}: Fit of the amplitude second order back-reaction $|\phi^{(2)c}_0|$ (dots) with the function $\sim e^{at^2+b}$ (red line), for $m=10^{-6}$. The fitting function is found to be in good agreement with the growing rate of $|\phi^{(2)c}_0|$. The $\phi$ background $\phi^{(0)}$ (green) is also presented in order to provide the timescale of the growth of $|\phi^{(2)c}_0|$ and to provide a scale of its relative amplitude.   \textsc{Bottom left}: Fitted value of $a(m)$ as a function of $m$ for $m=10^{-5.6}$ to $m=10^{-7}$. The fit (in red) shows that the points obey a linear relationship over the studied range of $m$.  \textsc{Bottom right}: Fitted value of $b(m)$ as a function of $m$, over the same range. The fit (in red) exibits that $b(m)$ grows as $\sqrt{m}$ over the studied values of $m$.}{fitexamplephi0}

\begin{figure}[] 
\includegraphics[scale=0.29]{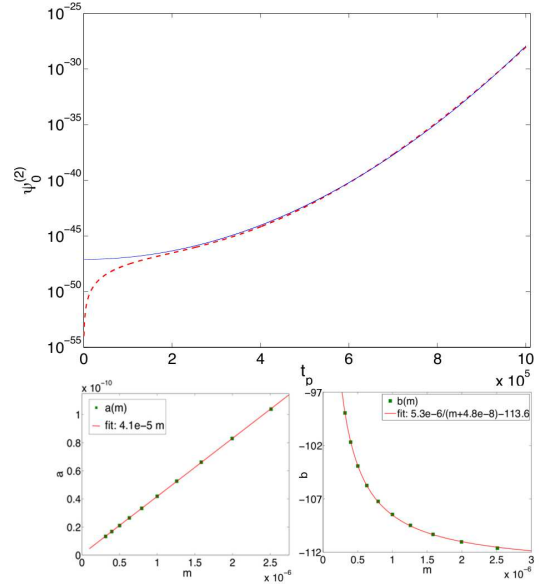}
\caption{\textsc{Top panel:} Fit of the amplitude of the second order back-reaction $|\psi^{(2)c}_0|$ (dashed red line) with the function $\sim e^{at^2+b}$ (blue line), for $m=10^{-6}$. The fitting function is found to be in good agreement with the growth rate of $|\psi^{(2)c}_0|$. \textsc{Bottom left}: Fitted value of $a(m)$ as a function of $m$ for values of $m$ ranging from $m=10^{-6.5}$ to $m=10^{-5.6}$. The fit (in red) shows that the points obey a linear relationship over the studied range of $m$.  \textsc{Bottom right}: Fitted value of $b(m)$ as a function of $m$, over the same range. The fit (in red) exhibits that $b(m)$ grows as $\frac{1}{m}$ over the studied values of $m$.}
 \label{fitexamplepsi0}
 \end{figure}

%\noindent between the growth of the back-reaction mode of the 
%two fields was to be expected, since in both cases it is the mass of the tachyonic 
%field that triggers the instability and that sets its timescale.

A fitting function very similar to the one found above for $|\phi_0^{(2),c}|$ is found to match the growth of $|\psi_0^{(2),c}|$ as $m$ is varied:
\begin{equation}
	\psi^{(2)c}_0 \, \approx \,  -exp\left[4.139e\mathrm{\textendash}5mt^2-\frac{5.336e\mathrm{\textendash}6}{(m+4.781e\mathrm{\textendash}8)}-111.6\right]
\end{equation}
This relation is displayed in Figure \ref{fitexamplepsi0} for the case $m=10^{-6}$, where it is found 
to be in agreement with the analytical values over a time scale of $m^{-1}$, but as explained above, it deviates over smaller time scales.

\subsection{Back-reaction on Long Wavelength $\phi$ Modes}

We now consider the case of main interest, i.e. the evolution of the large wavelength 
perturbation modes $\phi^{(2)c}_{\mathbbmtt{n}}$ (with $\mathbbmtt{n}$ small) at
second order in perturbation theory. Their evolution is governed by equation (\ref{2ndphiode}), 
with the interaction term replaced by (\ref{2ndphiode3d}). However, once simplified, the 
homogeneous version of these equations become exactly the homogeneous version of the 
equation we had for  $\phi^{(2)c}_0$, so that they can be solved by the same Green's function. 
Their solution thus reduces to the same integral as before, and only the prefactor will differ. The 
main difference is that the cancellation of the phases observed in the $\phi^{(2)}_0$ case is not 
present anymore, and sums over random phases need to be taken into account. The expectation 
value over phases must thus be performed as a random walk, as was done in the $\psi^{(2)}_0$ 
case. Terms with sums over 3 dimensions will again dominate over the ones with sums
over only 1 or 2 dimensions. Thus, we are left with the task of evaluating:
%%

%\begin{widetext}

\bea \label{2ndphiode3dsum}
%	&& - \frac{1}{8}\sum_{i,j,k=0}^{\frac{LH}{\pi}}\left[\psi^{(1)c}_{ijk}\psi^{(1)c}_{|i-n_x| |j-n_y| |k-n_z|} 
%	+\psi^{(1)c}_{ijk}\psi^{(1)c}_{(i+n_x) (j+n_y) (k+n_z)}\right.\nonumber \\
%	&&  \left. + 3\psi^{(1)c}_{ijk}\psi^{(1)c}_{|i-n_x| |j-n_y| (k+n_z)} 
%	+ 3\psi^{(1)c}_{ijk}\psi^{(1)c}_{|i-n_x| (j+n_y) (k+n_z)}\right] \\
	&& - \frac{1}{8}\sum_{i,j,k=0}^{\frac{LH}{\pi}}\left[\frac{(i^2+j^2+k^2)^{-1/4}}{\left[(i-n_x)^2+(j-n_y)^2+(k-n_z)^2\right]^{1/4}} \right. \nonumber \\
	&& + \frac{(i^2+j^2+k^2)^{-1/4}}{[(i+n_x)^2+(j+n_y)^2+(k+n_z)^2]^{1/4}} \nonumber \\
	&& + \frac{3(i^2+j^2+k^2)^{-1/4}}{[(i-n_x)^2+(j-n_y)^2+(k+n_z)^2]^{1/4}} \nonumber \\
	&&+ \left. \frac{3(i^2+j^2+k^2)^{-1/4}}{[(i-n_x)^2+(j+n_y)^2+(k+n_z)^2]^{1/4}}\right] 
\eea

%\end{widetext}

%\EPSFIG[scale=0.3595]{background_fluctuationsm6.eps}{\scriptsize  \textsc{Main graph}: Comparison on log-log scale of the amplitude growth of the $\phi$ back-reaction on large-scale fluctuations (blue) with the growth of the first order $\psi$ perturbations, for $\mathbbmtt{n}$ ranging from 10 to 10000 (red to yellow spectrum), and for $m=10^{-6}$. The evolution of the $\phi$ background is also shown to provide the overall timescale of the evolution. back-reaction becomes dominant within much less than one Planck time for all long wavelength modes considered. \textsc{Small graph}: Linear zoom-in of the back-reaction on long wavelength $\phi$ modes (blue) and first order fluctuations modes (red to yellow spectrum). $|\phi^{(2)c}_{\mathbbmtt{n}}|$ reach the amplitude of the studied long wavelength fluctuations within $\mathcal{O}(10^{-5})$.}{phiback-reactiontimeonmodes}

\noindent where each term can be summed independently as a distinct sum (again with
random phases). 
%%RB random phase comment added above

%%LPL slight clarification added below
To evaluate these, we assume, for the moment, that all modes are in phase. Under this assumption, 
we compute the first and second sums separately, and note that the third and fourth 
will be bounded by the values of the two firsts. Since they turn out to differ by a negligible amount, 
we will be able to set them all to be equal to each other. We then restore the randomness of the 
phases to compute the sums under the approximation that they are equal their monopole term in a 
multipole expansion.

Starting with the first sum, we note that it is bounded bellow by the sum (\ref{monopoleterm}), which 
is in fact the monopole contribution to its value in a multipole expansion. Since we are considering 
long wavelength modes ($\mathbbmtt{n}$ small), setting the sum equal to its monopole contribution 
constitutes a first approximation of its value. However, we wish to know how good this approximation 
will be as a function of $\mathbbmtt{n}$. To answer this question, we compute the difference 
between the sum and the monopole term (\ref{monopoleterm}), term by term:
%%

%\begin{widetext}

\bea
	&& \frac{\left(i^2+j^2+k^2\right)^{-1/2}}{\left(\frac{(i-n_x)^2+(j-n_y)^2+(k-n_z)^2}{i^2+j^2+k^2}\right)^{1/4}}-\left(i^2+j^2+k^2\right)^{-1/2} \nonumber \\
%	&=& \, \left(\frac{1-\left(1+\frac{-2(in_x+jn_y+kn_z)+n_x^2+n_y^2+n_z^2}{i^2+j^2+k^2}\right)^{1/4}}{\left[\left(i^2+j^2+k^2\right)\left((i-n_x)^2+(j-n_y)^2+(k-n_z)^2\right)\right]^{1/4}}\right) \nonumber \\
%	&\approx& \, \left(\frac{1-1+\frac{\frac{1}{2}(in_x+jn_y+kn_z)-\frac{1}{4}(n_x^2+n_y^2+n_z^2)}{i^2+j^2+k^2}}{\left[\left(i^2+j^2+k^2\right)\left((i-n_x)^2+(j-n_y)^2+(k-n_z)^2\right)\right]^{1/4}}\right)
	\nonumber \\
	&\leq& \, \frac{\frac{1}{2}(in_x+jn_y+kn_z)}{\left[i^2+j^2+k^2\right]^{5/4}}-\frac{\frac{1}{4}(\mathbbmtt{n}^2)}{\left[i^2+j^2+k^2\right]^{5/4}}
\eea

%\end{widetext}

This expression puts an upper bound on the term by term difference between the sum we want to 
evaluate and its monopole approximation (\ref{monopoleterm}). This difference can then be
summed over the three indices $i$, $j$, $k$, in order to obtain an upper bound on the total difference. 
To do so, we again use integrals to approximate the sums, and perform them numerically. 
We obtain, for $m=10^{-6}$ in Planck mass units:
\bea
	&& \frac{n_x+n_y+n_z}{2}\int_0^{\frac{LH}{\pi}}\frac{didjdk~(i)}{\left[i^2+j^2+k^2\right]^{5/4}} \nonumber \\
	&\approx& \,  \frac{n_x+n_y+n_z}{2}1.8\times 10^{35}
\eea
and
\begin{equation}
	-\frac{\mathbbmtt{n}^2}{4}\int_0^{\frac{LH}{\pi}}\frac{didjdk}{\left[i^2+j^2+k^2\right]^{5/4}} \, 
	\approx \, -\frac{\mathbbmtt{n}^2}{4}2.1\times10^{12} \, .
\end{equation}

%\EPSFIG[scale=0.38]{background_fluctuationsm6psi.eps}{\scriptsize \textsc{Main graph}: Comparison on log-log scale of the amplitude growth of the $\psi$ back-reaction on large-scale fluctuations (blue) with the growth of the first order $\psi$ perturbations, for $\mathbbmtt{n}$ ranging from 10 to 10000 (red to yellow spectrum), and for $m=10^{-6}$. The evolution of the $\phi$ background is also shown to provide the overall timescale of the evolution. back-reaction becomes dominant over less than one Planck time for all long wavelength modes considered. \textsc{Small graph}: Linear zoom-in of the back-reaction on long wavelength $\psi$ modes (blue) and first order fluctuations modes (red to yellow spectrum). $|\psi^{(2)c}_{\mathbbmtt{n}}|$ reach the amplitude of the studied long wavelength fluctuations within $\mathcal{O}(10^{-2})$.}{psiback-reactiontimeonmodes}

\begin{figure}[] 
\includegraphics[scale=0.38]{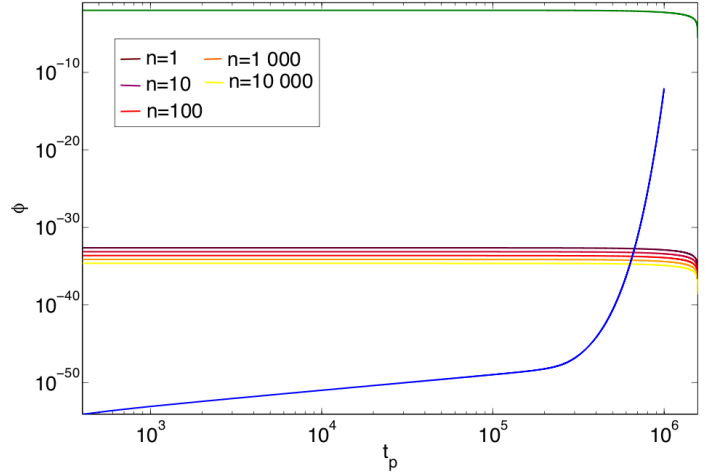}
\caption{Comparison on a log-log scale of the growth in amplitude of the $\phi$ back-reaction 
on large-scale fluctuations (blue) with the growth of the first order $\psi$ perturbations, for 
$\mathbbmtt{n}$ ranging from 1 to 10000 (red to yellow spectra), and for $m=10^{-6}$. 
The evolution of the $\phi$ background is also shown (topmost green curve) to provide the 
overall time scale of the evolution. Back-reaction becomes dominant within less than 
$mt=1$ for all long wavelength modes considered (about $5\times10^{5}$ Planck times).}
\label{phiback-reactiontimeonmodes}
\end{figure}

First, note that once the value of $\mathbbmtt{n}$ is fixed, which mode exactly we choose to 
consider on the momentum shell of radius $|\mathbbmtt{n}|$ will make at most a factor of 
$\sqrt{3}$ difference in the result we just obtained. Hence we can take 
$\mathbbmtt{n}=n_x$ and $n_y=n_z=0$ for simplicity.

Now, since the monopole term (\ref{monopoleterm}) is $\mathcal{O}(10^{49})$ for the 
%%RB minor change below and above (the number is adjusted)
parameter values which we are considering, we can be confident that the higher multipole 
corrections to the first sum are negligible as long as we choose 
$\mathcal{O}(\mathbbmtt{n})<10^{-12}$. Since we are interested in long wavelength modes, 
we can take the first sum in (\ref{2ndphiode3dsum}) to be equal to $\frac{ (Lm\phi_c)^2}{16\pi}$.

If we now tackle the evaluation of the second sum in (\ref{2ndphiode3dsum}), we find 
that (\ref{monopoleterm}) is an upper bound on its value. To compute the difference between 
this upper bound and the actual sum, we proceed by following the exact same 
approximation scheme as above. In doing so, we find that the difference is bounded by:
\begin{equation}
	\leq\sum_{i,j,k=0}^{\frac{LH}{\pi}}\left[\frac{\frac{i}{2}(n_x+n_y+n_z)}{\left[i^2+j^2+k^2\right]^{5/4}}+\frac{\frac{1}{4}(\mathbbmtt{n}^2)}{\left[i^2+j^2+k^2\right]^{5/4}}\right] \, ,
\end{equation}
which is once again negligible, provided that we choose $\mathcal{O}(\mathbbmtt{n})<10^{-12}$.

Hence, we conclude that the four sums in (\ref{2ndphiode3dsum}) can be taken to be 
equal without making any significant error. 
%%RB Remark about random phases added below
Now, performing the sums including random phases within this approximation,
we find that the evolution of the back-reaction term for long wavelength 
modes (with $\mathcal{O}(\mathbbmtt{n})<10^{-12}$) is  given by the solution:

\begin{figure}[] 
\includegraphics[scale=0.38]{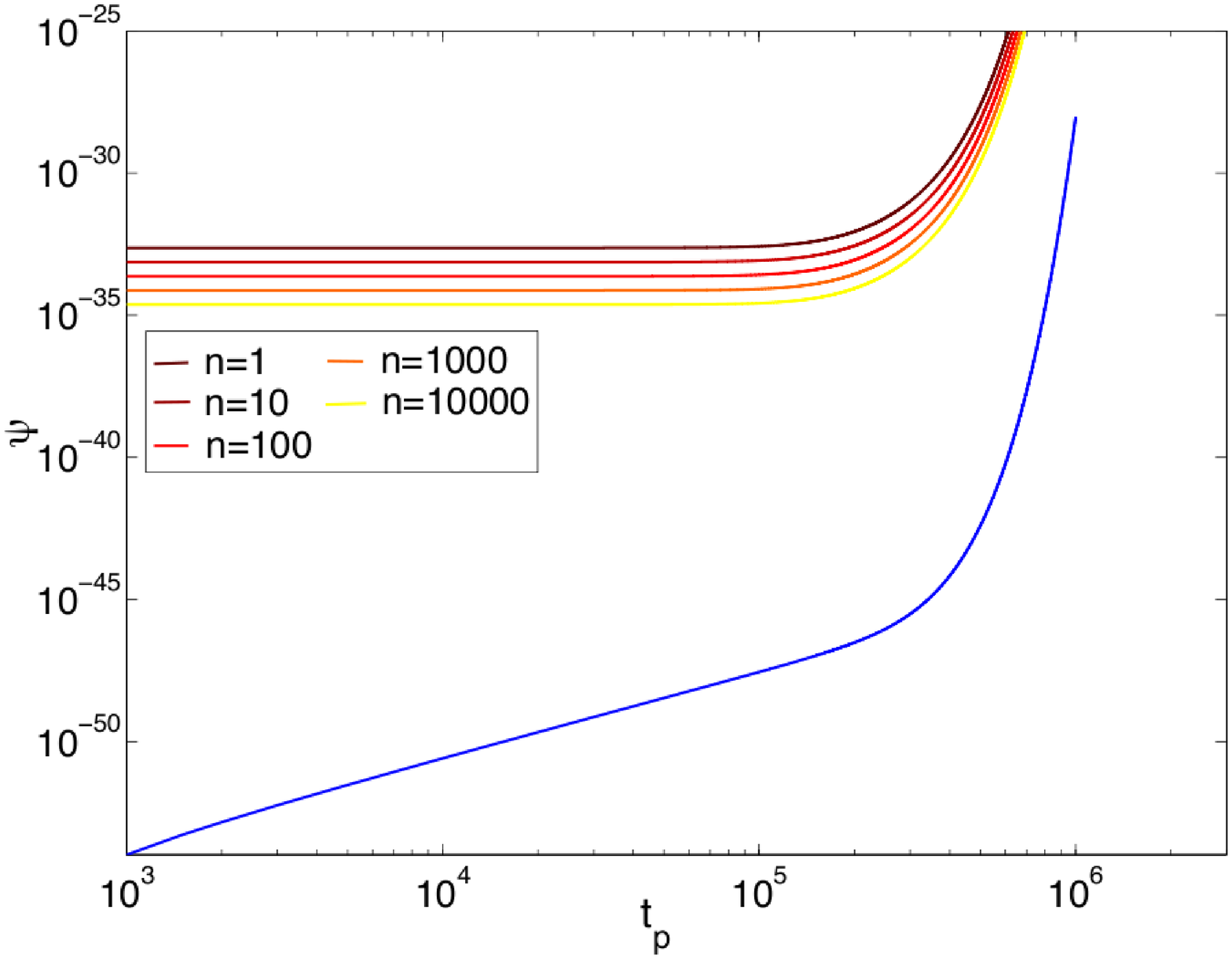}
\caption{Comparison on a log-log scale of the growth in amplitude of the $\psi$ back-reaction 
on large-scale fluctuations (blue) with the growth of the first order $\psi$ perturbations, 
for $\mathbbmtt{n}$ ranging from 1 to 10000 (red to yellow spectra), and for $m=10^{-6}$.  
The back-reaction never becomes dominant during the range of validity of our 
analytical approximations, that is, they do not become dominant before a time 
greater than $mt=3$.}
 \label{psiback-reactiontimeonmodes}
\end{figure}

\bea \label{phinsol}
	\phi^{(2)c}_{\mathbbmtt{n}}(t) \, &=& \, -\frac{g^2\phi_c v\lambda^{1/4}}{3^{1/4}8\pi^2mL^{3/2}}\int_0^{t}ds\sin(m(t-s)) \cos(ms) 
	\nonumber \\
	&& \, \times \cosh^2\left[\frac{g\phi_c(1-\cos ms)}{m}\right] \, .
\eea
%%
%%RB The above equation needs to be checked! LPL now it's fine
Since this integral form of the solution is the same as the one we obtained for $\phi^{(2)c}_0$, 
the growth and $m$ dependence of $\phi^{(2)c}_{\mathbbmtt{n}}$ (for long wavelength modes) 
will be exactly as studied above for the back-reaction on the background. The only difference comes in the overall normalisation of this integral, which is now greatly reduced by the random phase, instead of being proportional to the volume of first order excited modes in phase space. Our conclusions concerning the consistency and 
stability as a function of $m$ near the value $m=10^{-6}$ (in Planck units) thus still hold. 
In particular, we obtain the following fitting function:
\begin{equation}
\label{phinfittingfunction}
	\phi^{(2)c}_{\mathbbmtt{n}} \, \approx \,  -exp\left[8.071e\mathrm{\textendash}5 mt^2+\frac{7.477e\mathrm{\textendash}6}{(m+1.33e\mathrm{\textendash}8)}-146.4\right] \, .
\end{equation}
Note that for long wavelength modes, it is independent of $\mathbbmtt{n}$, and so the 
back-reaction term will have an identical growth for all modes with 
$\mathbbmtt{n}\sim\mathcal{O}(10^{12})$ .

Comparing the growth of the second order contributions to the amplitude of the long wavelength modes 
with the growth of the first order terms, we immediately see that, since the former grow as 
$\sim -e^{t^2}$, they will eventually come to dominate over the  corresponding first order terms. 
%%RB small adjustment of the text above
Indeed, the latter exhibit no instability since they simply oscillate as 
$\sim\cos mt$. The precise time scale on which the second order terms become dominant 
can be found by comparing the the fitting function (\ref{phinfittingfunction}) to obtain 
a semi-analytical estimate.

Figure \ref{phiback-reactiontimeonmodes} shows the evolution of long wavelength fluctuations 
and the corresponding back-reaction term for various values of $\mathbbmtt{n}$, for 
$m=10^{-6}$. For this value of $m$, it takes between $\mathcal{O}(10^{5})$ and $\mathcal{O}(10^{6})$ Planck times for the amplitude 
of the second order long wavelength mode to become equal to the first order fluctuation of that mode. This time scale is comparable to the longest typical time scales of the problem, $m^{-1}$, but is very long compared to the instability time scale in the tachyonic direction, $m_\psi$.

Since the only $\mathbbmtt{n}$ dependence of the intersection is through the normalization of the first 
order fluctuations modes in (\ref{sol1stphi}) (increasing by two orders of magnitude the momentum 
considered will reduce the amplitude of the first order perturbation by only one order of magnitude), 
the time when the second order effects start to dominate over the first order perturbations will depend 
only slightly on the value of $\mathbbmtt{n}$ considered, as long as only long wavelength modes 
are considered.
%%RB number above needs checking

%%LPL a agree with dropping the following paragraphs.
%%RB paragraph below dropped 
%This time scale found for the back-reaction for long wavelength fluctuations to become dominant 
%is much shorter than any other time scale present in the problem, i.e. it is shorter than both 
%$H^{-1}$ and the the times scales set by the inverse masses of the two fields. Specifically, it
%is shorter than the time scale of the tachyonic instability of the background waterfall field.

%When the second order terms start to dominate over the first order terms in the 
%fluctuations can be qualitatively understood by setting the initial expectation value for the 
%amplitude of a first order mode 
%equal to the volume in phase space of first order modes excited, times the amplitude of the 
%corresponding second order mode if it were driven by a single first order mode, and by finding 
%the time at which this equality holds. Taylor expanding to leading order in $t$ the integral in 
%(\ref{phinsol}) gives a rather good estimate of this latter quantity, so that the required time scale 
%is in fact given by the value of $t$ for which:
%%
%\begin{equation}
%	\frac{1}{4L}\frac{1}{(2\pi\mathbbmtt{n})^{1/2}} \, \sim \, %\left(\frac{ (L\lambda^{1/2}v^2)^2}{48\pi} \right)%\left(\frac{g^2\phi_c}{2\pi m}\frac{1}{(L)^2}\times \frac{mt^2}{2}\right)\, .
%	\frac{\lambda^{1/4} v}{10^2 L^{3/2}} m t^2 \, .
%\end{equation}
%%RB The above equation needs to be checked

\subsection{Back-Reaction on Long Wavelength $\psi$ Modes}

Repeating the above analysis to compute the back-reaction on the $\psi$
perturbations, i.e. on $\psi^{(2)c}_{\mathbbmtt{n}}$, for long wavelength modes we obtain
(once again taking into account that the phases are random):
%%RB Comment about random phases added above
%%
\bea \label{psinsol}
	\psi^{(2)c}_{\mathbbmtt{n}}(t) &=& 
	\frac{-g^2\phi_c\lambda^{\frac{1}{4}}v}{3^{\frac{1}{4}}\pi^2mL^{\frac{3}{2}}}\int_0^{t}ds \sinh\left[\frac{g\phi_c(1-\cos(m(t-s)))}{m}\right]  
	\nonumber \\
         && \, \times \cos^2(ms)\cosh\left[\frac{g\phi_c(1-\cos ms)}{m}\right]  \, ,
\eea
which is simply $8\psi^{(2)c}_0$. We thus obtain the following fitting function:
\begin{equation}
\label{psinfittingfunction}
	\psi^{(2)c}_{\mathbbmtt{n}}\, \approx \,  -exp\left[4.139e\mathrm{\textendash}5mt^2-\frac{5.336e\mathrm{\textendash}6}{(m+4.781e\mathrm{\textendash}8)}-113.7\right]
 \, ,
\end{equation}
which is still independent of $\mathbbmtt{n}$ for long wavelengths, i.e. for
$\mathbbmtt{n}\sim\mathcal{O}(10^{12})$ .

%%RB The paragraph below is changed substantially
Again, the growth of second order modes will be much faster than the growth of the corresponding 
first order modes, since their instability only grows as $\cosh(1-\cos mt)$.
However, the amplitude of the back-reaction term is suppressed by the random phases,
and so the time interval $\Delta t$ it takes before the second order term becomes more 
important than the first
order term for long wavelength modes is much larger than the time interval for the
onset of the instability of the first order fluctuations.
%so that the latter will be 
%outrun soon after $\phi$ crosses $\phi_c$. 
To find the exact crossing time, we need to find the 
intersection of (\ref{psinsol}) and (\ref{sol1stpsi}). The result, shown in 
Figure \ref{psiback-reactiontimeonmodes}, is $\mathcal{O}(10^{6})$ Planck times for $m=10^{-6}$.
This is much longer than the instability time scale for fluctuations of the waterfall field. 
Since once again the only $\mathbbmtt{n}$ dependence of this crossing point is through the 
normalization of the first order modes, the chosen value of the long wavelength mode $\mathbbmtt{n}$ 
will only mildly reduce this result as $\mathbbmtt{n}$ is increased. 

%The time scale $\Delta t$ found for the back-reaction on long wavelength fluctuations to become dominant is much shorter than any other time scale present in the problem. As above, this can be qualitatively understood as the time for which the initial amplitude of a first order mode equals  the volume in phase space of first order excited modes, times the amplitude of the corresponding second order mode if it were driven by a single first order mode. Taylor expanding the intagral in (\ref{psinsol}) to leading order in $t$ thus gives the approximate equation for $\Delta t$:
%%
%\begin{equation}
%	\frac{1}{(4L)^2}\frac{1}{\mathbbmtt{n}^{1/2}}=\left(\frac{4(Lm\phi_c)^2}{\pi} \right)\left(\frac{g^2\phi_c}{4\pi m}\frac{1}{(2L)^2}\times \frac{m^2t^3}{6}\right)\, .
%\end{equation}
%%
 
%Obtaining 
%$\Delta t$ shorter than $H^{-1}$ confirms that neglecting the expansion of space in the equations of  
%motion was consistent. 
Since $\Delta t$ is much larger than the inverse mass of the waterfall field $m_\psi$, 
we conclude that $\Delta t$ is sufficiently large so that the back-reaction cannot shut off the growth of the 
entropy fluctuations before they have time to induce an appreciable contribution to the
curvature perturbations during the early stages of the tachyonic instability. 
%%RB rest of the paragraph dropped
%One way of
%understanding this result is that the nonlinearities due to small-scale fluctuations will induce 
%a positive contribution to the square mass of the perturbations of the waterfall field early 
%enough to shut off the tachyonic resonance of the entropy mode before it becomes significant.
%Thus, it is consistent to focus on the primordial adiabatic fluctuations as the main source
%for curvature perturbations. 
%Had we obtained a large value for the crossing time, then the
%entropy fluctuations could have become so large that their contribution to the curvature
%perturbations  would have dominated. In that case, the computation of the spectrum
%of cosmological perturbations would have had to be re-considered, and there might
%even have arisen a contradiction with the small observed amplitude of curvature fluctuations 
%on large (cosmological) scales.

\section{Conclusions}

We have considered classical dynamics in a two field inflation model like hybrid inflation
in which one field - $\phi$ - is slowly rolling during inflation and a second field, the so-called
``waterfall field" $\psi$, develops a tachyonic instability once $\phi$ decreases below some
critical value $\phi_c$. We focus on the dynamics right after the instability develops,
which corresponds to the initial stages of preheating at the end of inflation.

%%RB text below changed
We have provided a perturbative analytical analysis of the back-reaction effects of linear
fluctuations on the background fields and on the perturbations themselves during the
initial phase of the tachyonic instability which arises at the end of a period of hybrid
inflation. We conclude that the tachyonic growth of the short wavelength fluctuation
modes leads - at second order in perturbation theory, the lowest order
at which mode mixing appears - to contributions which could in principle
shut off the tachyonic growth of long wavelength entropy modes.
However, in the case of random phases for the linear fluctuations considered
here, we find that time scale when the back-reaction terms become important
is much longer than the typical instability time scale of the tachyonic field.
This process cannot shut off the instability of entropy modes before the latter have had time
to become important. Note that there are other mechanisms which could dramatically
weaken the strength of the instability of the long wavelength modes of the waterfall
field, e.g. the decrease in amplitude of the linear fluctuations on super-Hubble
scales during inflation.

%%RB New paragraph
It is important to stress the crucial role which the assumption of random phases has
played in our analysis. Had we worked with correlated phases, then the back-reaction
effects would have been larger by a factor comparable to $L^{3/2}$ (in Planck units),
and back-reaction would then be able to cut off the tachyonic instability of modes
of the waterfall field long before they have time to develop.

%%RB Minor change
Our work complements the numerical
studies of this model \cite{Felder1,Felder2}. Our analytical approach has as a big
advantage that we can access the large dynamical range of scales which is
required in order to draw conclusions for modes of cosmological interest today
from dynamics taking place on microphysical scales at the end of inflation, the
scales which are studied in the numerical works. Disadvantages of our
approach based on a discrete Fourier space analysis are that it is purely
perturbative and that we cannot study
position space phenomena expected in the model such as the formation of
topological defects.

%%RB Minor adjustment
The reader may worry that the back-reaction effect which we find on the
adiabatic mode of the fluctuations, the $\phi$ fluctuations, might affect the
predicted curvature fluctuations of the model. However, this is not the case.
On scales larger than the Hubble radius, the curvature fluctuations are dominated
by gravity. Microphysical effects which occur at the time of reheating cannot
affect the adiabatic modes of the curvature perturbations. This follows in full
generality from the Traschen integral constraints \cite{Traschen} which hold
as long as we operate in the framework of General Relativity. In the case of
scalar field models of inflation, it was shown in \cite{FB1} that effects
at preheating cannot affect the adiabatic mode of the curvature fluctuations.

%%RB Minor adjustment below
The reader should also keep in mind that our analysis does not concern other
sources of entropy fluctuations in models like the
one considered here. Specifically, if topological defects are produced
during preheating, then the active evolution of the defects between the
time of formation and the present time will lead to a large contribution
to the entropy fluctuations. In fact, the specific model considered in this
paper, where both $\phi$ and $\psi$ are single component real fields,
would lead to domain walls and would in fact be ruled out because of the
domain wall problem \cite{DW}, the over-abundance of energy in the
walls. 

\begin{acknowledgments}

This project began as a summer student group project. We wish to thank
Eric Thewalt, and in particular Alexandra Tcheng and Gabriel-Dominique Marleau 
for their contributions to this study. The research of RB is supported in part by an 
NSERC Discovery Grant,
by the Canada Research Chairs Program and by a Killam Research Fellowship. 
The work of LPL is supported in part by NSERC Alexander Graham Bell CGS.

\end{acknowledgments}

\vskip0.3cm

\centerline{\bf Appendix: Sum over Random Phases}

In the evaluation of the sum of back-reaction terms, each term arises from an 
interaction between the two fields which each have a random phase and which, in general, 
do not cancel. We need to take the average over possible choices of phases not before 
studying the effects on back-reaction, but only after. That is, the sum present in the expression 
for the background fields and their fluctuations must be performed for every possible 
choice of phases, and only then the expectation value of the obtained displacement distribution 
must be taken.

Since we take the phases to be ramdom, this effect can be modeled by weighting randomly 
each amplitude, in the sum of the backreaction term, with 1 or -1. 
Calculating the expectation value of the absolute value of this sum is similar 
to calculating the average displacement of a random walk in one dimension, so we will call this 
quantity a random walk summation.      

For a one dimensional random walk with $N$ steps of size $\delta$, the expectation value of 
the displacement $(E|D|)$ is given by: 
\be
\lim_{N\rightarrow\infty}\frac{E|D|}{\sqrt N}= \delta\sqrt{\frac{2}{\pi}}
\ee

But in our case, this process is in fact slighly non-trivial, since the step size $\delta$ is a function of 
the radius ($r$) in momentum space: 
$\delta\rightarrow\delta(r)=\frac{1}{r}\frac{1}{2\pi(2L)^2}$. For simplicity, we consider the random walk 
sum on the different shells of the sphere in 
momentun space independently, and we give each shell a thickeness of 1. The 
number of steps is then $\frac{1}{8}$ the volume of the shell, that is:
\be
\sqrt{\frac{4\pi r^2}{8}}\underbrace{\Delta r}_{=1}
\ee 
As the random walk summation of one shell, we thus obtain:
\be
\sqrt{\frac{4\pi r^2}{8}}\frac{1}{r}\frac{1}{2\pi(2L)^2}\sqrt{\frac 2{\pi}}
=\sqrt{\frac{\pi}{2}}\frac{1}{2\pi(2L)^2}\sqrt{\frac 2{\pi}}
\ee
for each shell.  This result is now independent of the radius, which allows us to take a random walk 
over the shells, that is a random walk of $LH/\pi$ steps with stepsize $\frac{1}{2\pi (2L)^2}$ (note 
that just summing the above result over the shells is not the right way of proceeding, because the 
value calculated above could be $+$ or $-$ from one shell to the other, and doing so would thus 
greatly over-estimate the average result). We finally obtain:
\be
\sqrt{\frac{LH}{\pi}}\frac{1}{2\pi(2L)^2}\sqrt{\frac 2{\pi}}=\frac{\sqrt{2H}}{8\pi^2L^{3/2}}
\ee
We also did a root mean square approximation to what the expectation value of the absolute 
value of the sum is, and we got the same expression within a numerical factor of order 1.

%}
%\end{multicols}

\end{document}